\documentclass[prb,aps,twocolumn]{revtex4}
\usepackage{amsmath}
\usepackage{amssymb}
\usepackage {graphicx}

\newcommand{\beq}{\begin{equation}}
\newcommand{\eeq}{\end{equation}}
\newcommand{\bea}{\begin{eqnarray}}
\newcommand{\eea}{\end{eqnarray}}
\newcommand{\be}{\begin{equation}}
\newcommand{\ee}{\end{equation}}
\newcommand{\D}{\Delta}
\newcommand{\e}{\varepsilon}

\begin{document}

\title{$s+is$ State with Broken Time Reversal Symmetry in Fe-Based Superconductors}
\author{Saurabh Maiti, Andrey V. Chubukov}
\affiliation {~Department of Physics, University of Wisconsin,
Madison, Wisconsin 53706, USA}
\date{\today}

\begin{abstract}
We analyze the evolution of the superconducting gap structure in
strongly hole doped Ba$_{1-x}$K$_x$Fe$_2$As$_2$ between $x=1$  and
$x \sim 0.4$ (optimal doping). In the latter case, the pairing
state is most likely $s\pm$, with  different gap signs on hole and
electron pockets, but with the same signs of the gap on the two
$\Gamma$-centered hole pockets (a ++ state on hole pockets). In a
pure KFe$_2$As$_2$ ($x=1$), which has only hole pockets, laser
ARPES data suggested another $s\pm$ state, in which the gap
changes sign between hole pockets (a $+-$ state). We analyze how
++ gap transforms into a $+-$ gap as $x \to 1$. We found that this
transformation occurs via an intermediate $s+is$, state in which
the gaps on the two hole pockets differ in phase by $\phi$, which
gradually involves from $\phi = \pi$ (the $+-$ state) to $\phi =0$
(the ++ state). This state breaks time-reversal symmetry and has
huge potential for applications. We compute the dispersion of
collective excitations and show that two different Leggett-type
phase modes soften at the two end points of TRSB state.
\end{abstract}

\maketitle

\section{Introduction}

The high interest in iron based superconductors (FeSC) is
primarily due to two key reasons. The first is a hope that the
analysis of FeSCs will not only resolve the pairing mechanism in
these systems but also provide important insights into the
electronic pairing in a generic high-T$_c$ superconductor. The
second is a hope to explore multi-band structure of FeSCs and
discover novel exotic superconducting states which have not been
observed in other systems. Out of such novel superconducting
states, the most searched for are the ones which break
time-reversal symmetry. A spin-triplet time-reversal symmetry
broken (TRSB) $p_x \pm i p_y$ state has likely been found in
$Sr_2RuO_4$ \cite{Mackenzie}; the spin-singlet $d+id$ TRSB state
has not yet been observed experimentally, although it was once
proposed as a candidate state for high $T_c$ cuprate
superconductors \cite{laughlin}, and was recently predicted
theoretically to occur for fermions on  hexagonal and honeycomb
lattices near van-Hove doping~\cite{d+id}.

Several groups already searched for TRSB state in FeSCs by
exploring the idea that at least in  some FeSCs both $s-$wave and
$d-$wave channels are attractive
\cite{CW,Kuroki,theory1,DHL,RT1,RT3,RT2,hanke,khodas,spid3}, and
that one can, in principle, transform from $s-$wave to $d-$wave
pairing by varying system parameters -- electron~\cite{hanke} or
hole~\cite{RT3} doping, hybridization between electron
pockets~\cite{khodas}, or degree of magnetic
scattering~\cite{spid3}. In between, there is a co-existence
regime in which both $s$ and $d$ order parameters are present,
with relative phase $\pm \frac{\pi}{2}$, i.e., the system develops
a TRSB  $s\pm id$ superconductivity. The majority of proposals for
$s+id$ state are for electron-doped FeSCs, but up to now a d-wave
superconductivity has not been found in strongly electron-doped
Ba(Fe$_{1-x}$Co$_x$)$_2$As$_2$ nor in KFe$_2$Se$_2$-type systems
which contain only electron pockets.

In this communication, we discuss another possible realization of
TRSB state in FeSCs -- a purely $s-$wave state with phase
difference $\phi$ between superconducting order parameters on
different Fermi pockets, which is not a multiple of $\pi$. The
free energy of such a state is symmetric with respect to $\phi \to
- \phi$. This Z$_2$ symmetry (which corresponds to time reversal
since $\phi \to -\phi$ implies $\Delta \to \Delta^*$) is broken
when the system spontaneously chooses $\phi$ or $-\phi$. We label
such a state as $s+is$. The $s+is$ state has been discussed in
Refs~\onlinecite{zlatko,agtenberg,ng,tanaka,wang,japan,babaev,gs,stanev}
as a generic possibility of the superconducting order in the case
when there are more than two Fermi pockets and as a surface state
in a two-band superconductor~\cite{bobkov}. We show below that
TRSB $s+is$ state with varying $\phi$ can be realized in strongly
hole-doped Ba$_{1-x}$K$_x$Fe$_2$As$_2$ near $x =1$.

We begin by listing several facts about
Ba$_{1-x}$K$_x$Fe$_2$As$_2$. (i) Near optimal doping, $x \sim
0.4$,  ARPES \cite{ARPES1,ARPES2}, neutron
scattering\cite{lumsden}, penetration depth~\cite{Khasanov} and
 thermal conductivity\cite{more optimal,more optimal_1}
measurements give strong evidence for nodeless, near-constant
$s\pm$ gap, which changes sign between hole and electron pockets.
This is consistent with theoretical
calculations\cite{theory1,Kuroki,DHL,RT1,LAHA,s-wave}.
 (ii) Recent measurements on Ba$_{1-x}$K$_x$Fe$_2$As$_2$
with $x=1$ (Refs. \onlinecite{Ding,ARPES}) and $x =0.93$ and $x =0.88$ (Ref. \onlinecite{shin_last}) indicate that superconducting $T_c$ most likely
remains non-zero from $x=0.4 \to 1$.
 (iii) For the $x=1$ material KFe$_2$As$_2$, ARPES measurements
\cite{Ding,ARPES} show that only hole pockets are present.
According to theory, in this situation, both $d-$wave and $s-$wave
pairing amplitudes are
attractive\cite{RT3,RT2,Kuroki,LAHA,korshunov}, and which state
wins depends on delicate interplay between system parameters.
$d-$wave gap is the largest on the hole pocket, which in the
unfolded Brillouin zone is centered at $(\pi,\pi)$
(Refs.\onlinecite{RT2,LAHA}), and $s-$wave gap is the largest on
the two $\Gamma-$centered hole pockets (GCP's), and changes sign
between them~\cite{korshunov}. The existing experiments point to
either $d-$wave and $s-$wave gap symmetry: thermal
conductivity~\cite{LT,Wang_1} and specific heat~\cite{spec} data
on KFe$_2$As$_2$ have been interpreted in favor of $d-$wave gap
symmetry, while laser ARPES measurements \cite{ARPES} and other
thermal conductivity data \cite{kyoto_thermal} have been
interpreted as evidence for $s-$wave.

If the gap in KFe$_2$As$_2$ is $d-$wave, one should obviously
expect a transition from $d-$wave to $s\pm$ state in
Ba$_{1-x}$K$_x$Fe$_2$As$_2$ as $x$ decreases from 1, and the
region of an intermediate $s+id$ state at low $T$ \cite{RT3}. In
this work we consider what happens if the gap in KFe$_2$As$_2$ is
$s-$wave. At a first glance, one might expect a gradual evolution
of the gap structure with $x$ as the symmetry at $x=1$ is the same
as at optimal doping. On a more careful look, however, we note
that at optimal doping the gaps on the two GCP's have equal signs
(a ++ state), while in $s-$wave state of KFe$_2$As$_2$ they are of
opposite signs (a $+-$ state). The issue then is how a $+-$ gap
transforms into a $++$ gap between $x=1$ and optimal doping. We
show that this transformation occurs via an intermediate $s+is$
state in which the relative phase $\phi$ of the superconducting
order parameters on the two GCP's gradually evolves between $\pi$
(the $+-$ state) and $0$ (the ++ state). The system spontaneously
chooses either clock-wise or counter-clockwise evolution (i.e.,
positive or negative $\phi$) and by this breaks time-reversal
symmetry.

To illustrate the emergence of the $s+is$ state we first consider
in Sec. 2 the minimal model with two identical GCP's and two
electron pockets, all with the same density of states $N_0$, and
with the two angle-independent repulsive interactions -- $U_{hh}$
between the two GCP's and  $U_{he}$ between hole and electron
pockets. A three-band version of this model has been has been
considered in Refs.
\onlinecite{zlatko,tanaka,wang,japan,babaev,stanev}). The
interaction $U_{hh}$ gives rise to $+-$ gaps on the two GCP's,
while $U_{he}$ gives rise to an $s\pm$ state with different signs
of the gaps on the two hole pockets.  We model the doping
dependence by varying the strength of hole-electron coupling
$U_{he}$ and analyze the system evolution with $U_{he}/U_{hh}$. We
show that it occurs via a TRSB state. In Sec. 3 we extend the
model and include intra-pocket repulsions and anisotropy between
the two hole pockets. We show that the TRSB state still exists in
a certain parameter range, but for non-equivalent hole pockets the
region of TRSB state is separated from $T_c$ line.  We present our
conclusions in Sec. 4. Technical details of our analysis are
presented in Appendixes A-C.  In Appendix C we also discuss
plasmon mode in a clean 3D superconductor.

\section{TRSB in the minimal model}

The Hamiltonian of the minimal model is~\cite{AVC_RG}
$H=H_{kin}+H_{int}$, where $H_{kin}= \sum_{i,k,\alpha} \e_k
(c^{\dag}_{i k\alpha}c_{i k\alpha} - f^{\dag}_{i k\alpha}f_{i
k\alpha})$ and $H_{int}=\frac{1}{2} \sum_{k,\alpha,\beta}
\left[U_{hh} b_{c_1k}^{\dag}b_{c_2 k} + \sum_{i,j} U_{he}
b_{c_ik}^{\dag}b_{f_i k} + h.c \right]$ where $b_{x\,k}= \sum_k
x_{k \uparrow \alpha}x_{k \downarrow \beta}$ and $x\in
\{c_1,c_2,f_1,f_2\}$; and $i,j=1,2$ number the hole pockets ($c$)
and electron pockets ($f$). We define superconducting gaps on two
hole pockets as $\Delta_{h_1}$ and $\Delta_{h_2}$ and the gap on
electron pockets as $\Delta_{e_1}$ and $\Delta_{e_2}$.  We neglect
the angular dependence of $U_{he}$ in which case $\Delta_{e_1} =
\Delta_{e_2}$ because $U_{he}$ for pockets $e_1$ and $e_2$ are
equivalent due to C$_4$ symmetry of the underlying lattice. The
equivalence between $\Delta_{e_1}$ and $\Delta_{e_2}$ persists
even if we include intra-pocket interactions and inter-pocket
interaction between the two electron pockets.

The set of linearized equations for $\Delta_{h_1}$,
$\Delta_{h_2}$, and $\Delta_{e_1} = \Delta_{e_2} = \Delta_e$ is
obtained straightforwardly and reads \bea \label{eq:lin gap
eq}\left(
\begin{array}{c}
\D_{h_1}\\
\D_{h_2}\\
\D_{e}
\end{array}\right)&=&
-L \left(
\begin{array}{cccc}
0&u_{hh}&2u_{he}\\
u_{hh}&0&2u_{he}\\
u_{he}&u_{he}&0
\end{array}\right)
\left(
\begin{array}{c}
\D_{h_1}\\
\D_{h_2}\\
\D_{e}
\end{array}\right)
\eea where $u_{he} = U_{he} N_0, u_{hh} = U_{hh} N_0$, $N_0$ is
the density of states, $L\equiv ln \left(
\frac{2\Lambda}{T_c}\right)$, and $\Lambda$ is the upper cut-off
for the pairing.  This set can be easily solved. For $u_{he}>
u_{hh}/\sqrt{2}$, the eigenfunction with the largest eigenvalue is
the $++$ solution $(1,1, -\gamma)$, where $\gamma =
\frac{u_{hh}}{4u_{he}}+\sqrt{1+\left( \frac{u_{hh}}{4u_{he}}
\right)^2}$, and for $u_{he} < u_{hh}/\sqrt{2}$, is a $+-$
solution $(1,-1,0)$. Precisely at $u_{he}= u_{hh}/\sqrt{2}$ the
two states become degenerate and  $a (1,1, -\gamma) +b (1,-1,0)$
with arbitrary ratio of $a/b$ becomes an eigenfunction. To see
what happens immediately below $T_c$ at this critical
$u_{he}/u_{hh}$ we expand the Free energy in powers of
$\Delta_{h_i}$ and $\Delta_{e_i}$ to fourth order and
obtain (see Appendix A)

\bea \label{eq:free1} \mathcal{F}&=& \mathcal{F}_0 -K_0 \left(
|a|^2+|b|^2\right) + K_1 \left( |a|^2+|b|^2\right)^2 \nonumber \\
&&+ K_2|a^2+b^2|^2 + K_3 |a|^4 \eea where $K_0 \propto T_c -T$,
$K_{1,2}>0$, and  $0>K_3 > -2K_2$.
 Minimizing with respect to $a$ and $b$ we immediately obtain $b =
\pm i a\sqrt{1+\frac{K_3}{2K_2}}$, i.e., the ++ and $+-$ states
co-exist with relative phase $\pm\pi/2$. As a consequence,
immediately below the degeneracy point, the system selects an
$s+is$ state, which breaks time reversal symmetry (a TRSB state).

\begin{figure*}[htp]
$
\begin{array}{cc}
\includegraphics[width=0.9\columnwidth]{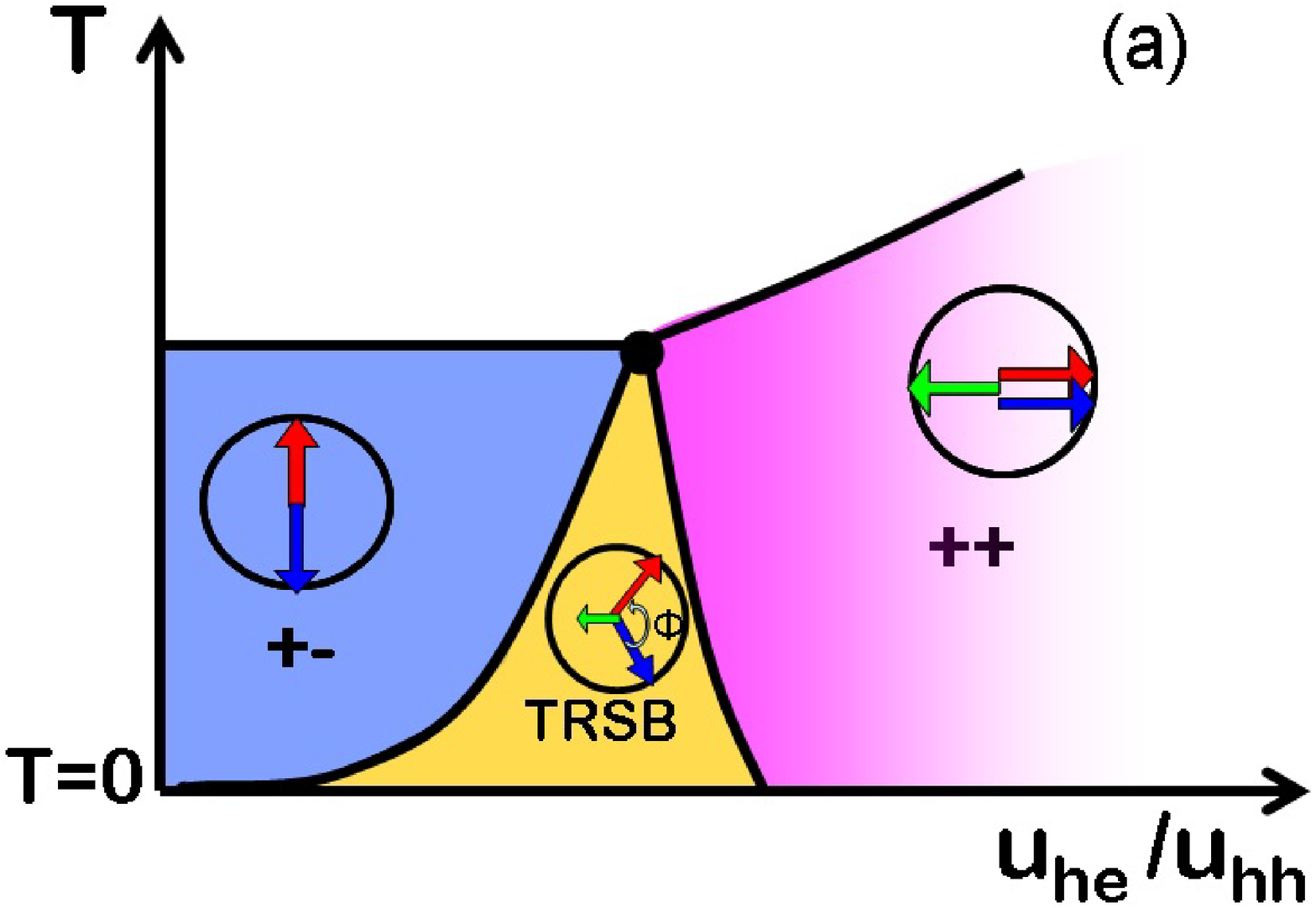}&
\includegraphics[width=0.9\columnwidth]{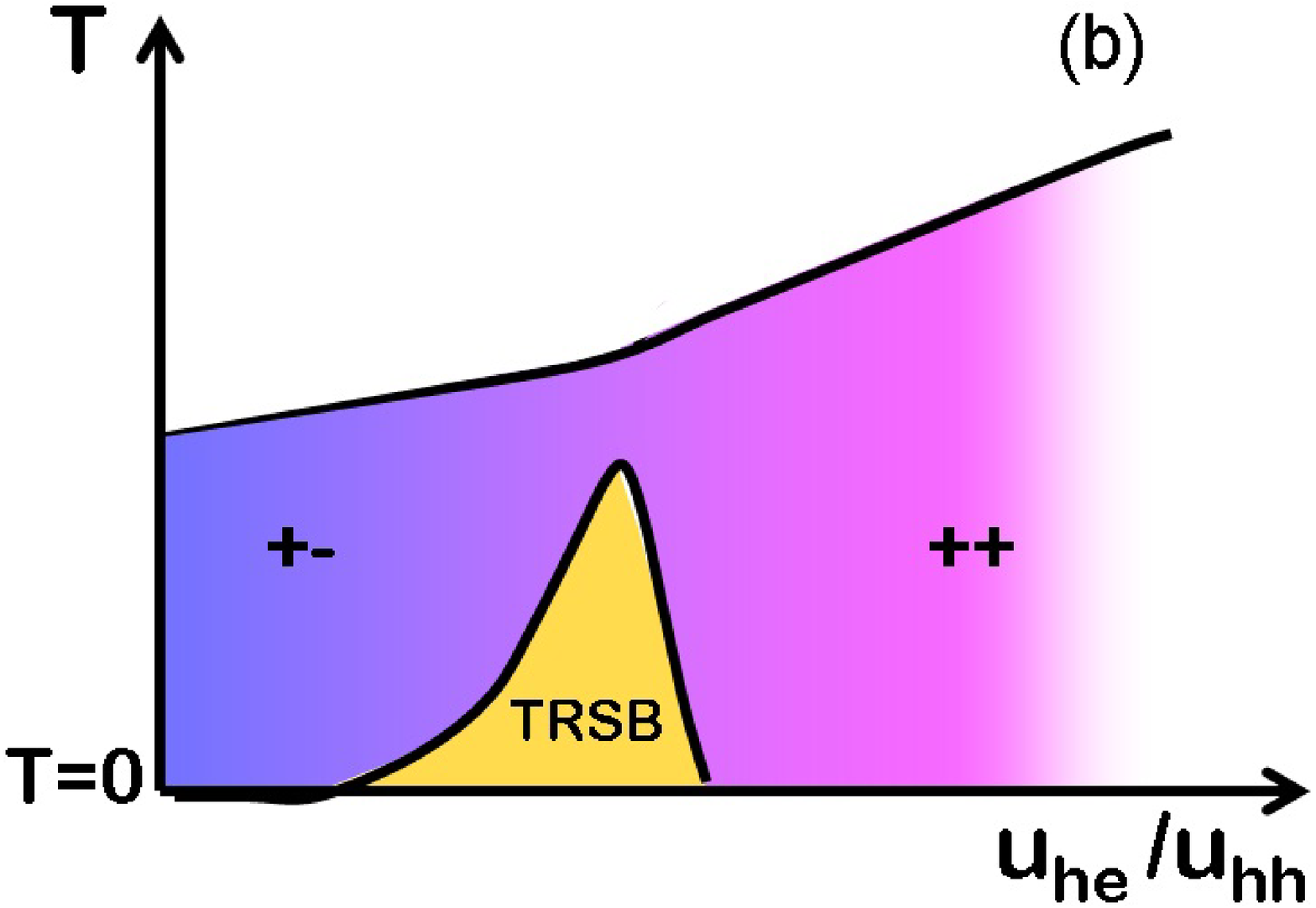}
\end{array}
$
\caption{\label{fig:trsb_schematic}Qualitative phase diagram for
Ba$_{1-x}$K$_x$Fe$_2$As$_2$ at $x \leq 1$. We model the doping
dependence by varying the ratio of inter-pocket electron-hole and
hole-hole interactions $u_{he}/u_{hh}$ which roughly scales as
$1-x$. The $+-$ state has gaps of opposite signs on the two GCP's
and no gap on electron pockets, the $++$ state is an ordinary
$s\pm$ state in which the gaps have opposite signs on hole and
electron pockets, and between them is the TRSB state.
The gap structures are pictorially presented inside each region by vectors placed inside the circles.
 The magnitudes of the vectors represent $|\Delta_i|$ and the angles
represent the  phases.
Cases $(a)$ and $(b)$ are for equal and non-equal intra-pocket
interactions ($u_{h_1}$ and $u_{h_2}$) for the two hole pockets,
respectively . For ($a$), the TRSB state starts right at $T_c$ and
extends into a finite range at $T=0$. For ($b$), the TRSB region
splits off from the $T_c$ line and is only accessible at lower
temperatures, while immediately below $T_c$ the $+-$ state
gradually evolves into the $++$ state as $u_{he}/u_{hh}$
increases.}
\end{figure*}

\begin{figure}[htp]
\includegraphics[width=0.9\columnwidth]{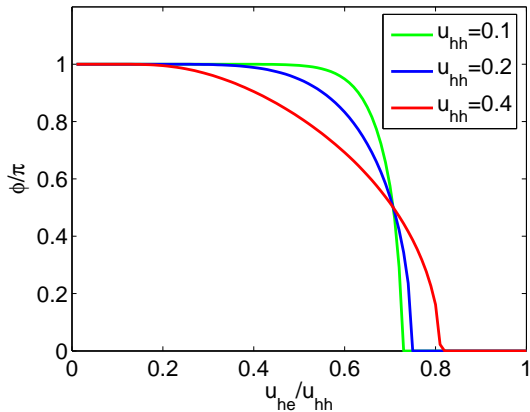}
\caption{\label{fig:phase} Variation of the relative phase $\phi$
of the gaps on two hole pockets with $u_{he}$. This phase is zero
for $u_{he}>u_{he}^{max}$, but becomes non-zero at smaller
$u_{he}$ and eventually reaches  $\phi=\pm \pi$ at $u_{he} =0$.
When $|\phi|$ is between $0$ and $\pi$, it can be either positive
or negative, and the choice breaks $Z_2$ time-reversal symmetry.
The width of the TRSB region is controlled by inter-pocket
hole-hole interaction $u_{hh}$ and increases when $u_{hh}$ gets
larger.}
\end{figure}

Inside the TRSB state we can set $\Delta_e$ to be real and
$\Delta_{h_1} = \Delta e^{i\phi/2}, \Delta_{h_2} = \Delta
e^{-i\phi/2}$. We solved the set of three non-linear gap equations
at $T=0$ (see Appendix B) and found that TRSB state exists between
$u^{min}_{he} =0$ and $u^{max}_{he} \approx
\frac{u_{hh}}{\sqrt{2}} \left(1 + \frac{u_{hh}}{4}\log{2}\right)$.
At the lower boundary, the TRSB state borders $+-$ state and the
relative phase reaches $\phi =\pi$, at the upper boundary the TRSB
state borders $++$ state and $\phi =0$. In between, \bea &&\phi =
\pm 2 \arccos\left[\frac{u_{hh}}{2u_{he}} e^{(2u^2_{he}
-u^2_{hh})/(2 u^2_{he} u_{hh})}\right] \label{ch_1} \eea We show
the evolution of the relative phase $\phi$ on the two hole pockets
with $u_{he}/u_{hh}$  in Fig. \ref{fig:phase}

Combining the results at $T_c$ and at $T=0$, we obtain the phase
diagram shown in Fig. \ref{fig:trsb_schematic} (a). The TRSB state
exists in the `triangle' which begins as a point at $T_c$ and
extends to a finite interval at $T=0$.

\subsection{Collective modes}

The existence of phase transitions at the
boundaries of the TRSB state implies that there must be soft
collective excitations. In a generic multi-gap superconductor
there are three types of collective excitations: (i) variation of
the overall phase, (ii) variations of relative phases of different
gaps (Leggett modes~\cite{leggett}), and (iii) variations of the
gap magnitudes. The overall phase mode is coupled by long-range
Coulomb repulsion to density variations and becomes a
plasmon~\cite{cote_griffin,dirty}. The other modes do not couple
to density variations are generally either overdamped or have
energy close to $2\Delta$. However, near the boundaries of the
TRSB state, some of these modes soften.

\begin{figure}[t]
\includegraphics[width=0.9\columnwidth]{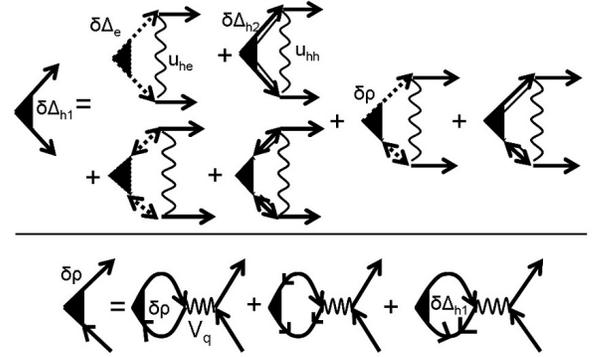}
\caption{\label{fig:fig2a}The diagrammatic representation of the
equations for dispersion of collective modes. The equations for
other $\delta \Delta_j$ are similar to the one for $\delta
\Delta_{h_1}$ and are not shown. Wavy lines - interactions
$u_{ij}$, chain-saw line -- Coulomb interaction $V_q$. The bare
vertices are not shown. }
\end{figure}

We analyzed the dispersion of collective excitations in our model
by introducing small perturbations in the form of pairing and
density vertices with non-zero external momentum and frequency
($\delta \Delta_{h_1}, \delta \Delta_{h_2}, \delta \Delta_e$, and
$\delta \rho_i$, $i=1,2,3$) and calculating the fully renormalized
vertices (see Fig. \ref{fig:fig2a}). Each $\delta \Delta$ is
generally a complex function $\delta \Delta_i = \delta^R_i + i
\delta^I_i$, so for arbitrary momentum $q$, the problem reduces to
solving the set of nine coupled equations for $\delta^R_i$,
$\delta^I_i$, and $\delta \rho_i$. We verified, however,  that at
small $q$,  when short-range interactions $u_{hh}, u_{he}$ can be
neglected compared to the static Coulomb interaction $V(q)$, all
three $\delta \rho_i$ are equivalent, because the Coulomb
repulsion does not distinguish between the different fermions
(Refs. \onlinecite{leggett,LR1}). In this approximation, i.e.,
$\delta \rho_i = \delta \rho$, and the number of equations reduces
to seven.

The equation for $\delta \Delta_{h_1}$ is graphically shown in
Fig. \ref{fig:fig2a}. Other equations are similar. In explicit
form we have
\begin{widetext}
\bea\label{eq:filler} 2\delta_i^R &=& 2\delta_i^R(0)+\sum_j
u_{i,j}\left[\Pi^{11}_{jj}\delta^R_j-\Pi_{jj}^{12}\delta^I_j+\Pi^{13}_{jj}\delta\rho_j\right]\nonumber\\
-2\delta_i^I &=& -2\delta_i^I(0)+\sum_j
u_{i,j}\left[\Pi^{21}_{jj}\delta^R_j-\Pi_{jj}^{22}\delta^I_j+\Pi^{23}_{jj}\delta\rho_j\right]\nonumber\\
2\delta\rho_i &=& \sum_j
2N_0V(q)\left[\Pi^{31}_{jj}\delta^R_j-\Pi_{jj}^{32}\delta^I_j+\Pi^{33}_{jj}\delta\rho_j\right]\nonumber\\
\eea
\end{widetext}
where $\delta(0)$ are bare pairing and density vertices which we
introduced as small corrections to the Hamiltonian (see Appendix
C), $V(q)$ is long-range Coulomb potential, and the components of
the matrix $u_{ij}$ are

\bea \label{eq:U} u_{i,j}  = \left(
\begin{array}{ccc}
0&u_{hh}&2u_{he}\\
u_{hh}&0&2u_{he}\\
u_{he}&u_{he}&0\\
\end{array}
\right) \eea
Further,
 $\Pi^{ab}_{ii} = \Pi^{ab}_{ii} (q, \Omega) = \frac{1}{N_0}\int d^2
k d \omega/(2\pi)^3 Tr\left[ \mathcal{G}_i (k, \omega)\sigma^a
\mathcal{G}_i (k+q, \omega + \Omega)\sigma^b\right]$, where
$\sigma^a$ are Pauli matrices, $\mathcal{G}_i$ are Nambu
Green's function of a superconductor, $i\in\{c_1,c_2,e\}$.

In explicit form we have (see Appendix C for details)
\bea\label{eq:pi_full} \Pi_{ii}^{11}(\vec{q},\Omega)&=&-\left[
2L_i-1-\cos\phi-\left(\frac{4}{3}-
\frac{2}{3}\cos\phi\right)X^2_i\right]\nonumber\\
\Pi_{ii}^{22}(\vec{q},\Omega)&=&-\left[
2L_i-1+\cos\phi-\left(\frac{4}{3}+
\frac{2}{3}\cos\phi\right)X^2_i\right]\nonumber\\
\Pi_{ii}^{33}(\vec{q},\Omega)&=&-\left[
2-\frac{4}{3}\left(\frac{\Omega}{2\Delta_i}
\right)^2\right]\nonumber\\
\Pi_{ii}^{12}(\vec{q},\Omega)&=&-\sin\phi\left[
1-\frac{2}{3}X^2_i\right]\nonumber\\
&=&\Pi_{ii}^{21}(\vec{q},\Omega)\nonumber\\
\Pi_{ii}^{13}(\vec{q},\Omega)&=&-\,\frac{i\Omega}{\Delta_i}\sin\frac{\phi}{2}\left[
1-\frac{2}{3}X^2_i\right]\nonumber\\
&=&-\Pi_{ii}^{31}(\vec{q},\Omega)\nonumber\\
\Pi_{ii}^{23}(\vec{q},\Omega)&=&-\,\frac{i\Omega}{\Delta_i}\cos\frac{\phi}{2}\left[
1-\frac{2}{3}X^2_i\right]\nonumber\\
&=&-\Pi_{ii}^{32}(\vec{q},\Omega)
\eea
where $L_i =\ln\left(\frac{2\Lambda}{\Delta_i}\right)$ and
$X^2_i=-\left(\frac{\Omega}{2\Delta_i}\right)^2+\frac{v_F^2}{2}
\left(\frac{q}{2\Delta_i}\right)^2$.

\begin{figure}[t]
\includegraphics[width=0.9\columnwidth]{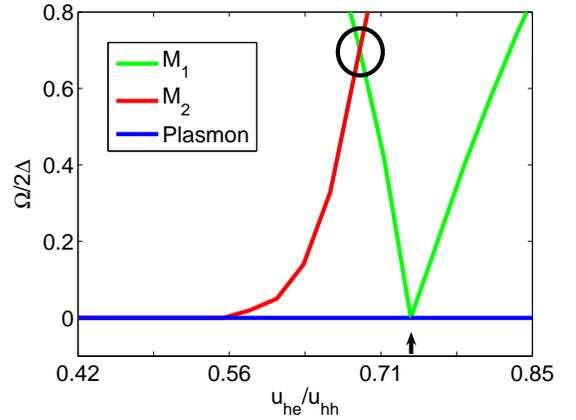}
\caption{\label{fig:fig2b} Doping evolution of the frequencies of
the relevant collective modes at $q=0$. The plasmon mode frequency
vanishes at $q=0$ for all $u_{he}$. Mode $M_1$ describes relative
phase fluctuation of the two hole gaps. It
softens\cite{stanev,japan,babaev} at the transition point between
++ and TRSB states (at $u_{he}=u_{he}^{max}$, indicated by the
arrow). Mode $M_2$ describes  coupled antisymmetric phase
fluctuation of the two hole gaps and longitudinal fluctuation of
the electron gap. This mode lies below twice the energy of the
electron gap and softens at the boundary between TRSB and $+-$
states, at $u^{min}_{he}=0$.  Numerically, the energy of the $M_2$
mode becomes small already for $u_{he} \leq u_{he}^{max}$ because
electron gap rapidly decreases with decreasing $u_{he}$. The
circle represents the case discussed in Ref.~\onlinecite{stanev}.}
\end{figure}

It is intuitive to reexpress Eq. \ref{eq:filler} as
\bea\label{eq:reexpress}
&&2\sum_j(u^{-1})_{ij}\delta^{a}_j=2\sum_j(u^{-1})_{ij}\delta^{a}_j(0)+\sum_b\Pi^{a,b}_{ii}\delta^{b}_i\nonumber\\
&&2\delta\rho = -\sum_j
2N_0V(q)\left[\Pi^{31}_{jj}\delta^R_j-\Pi_{jj}^{32}\delta^I_j+\Pi^{33}_{jj}\delta\rho\right]\eea
where $\delta^b_i=(\delta^R_i,-\delta^I_i,\delta\rho)$. This
7$\times$7 set can be cast into the form \be {\underline K} (q,
\Omega) ~{\vec { \delta}} = {\underline u}^{-1} ~ {\vec \delta
(0)} \label{ch_4} \ee

where ${\vec {\delta}}$ is a 7-component vector with elements
${\delta}^R_i, -{ \delta}^I_i, {\delta}_\rho$ ($\vec{\delta}(0)$ is
the bare vertex),
\begin{widetext}
\bea \label{eq:u} \underline{u}^{-1} = \left(
\begin{array}{ccccccc}
\frac{1}{u_{hh}}&-\frac{1}{u_{hh}}&-\frac{1}{u_{he}}&0&0&0&0\\
-\frac{1}{u_{hh}}&\frac{1}{u_{hh}}&-\frac{1}{u_{he}}&0&0&0&0\\
-\frac{1}{2u_{he}}&-\frac{1}{2u_{he}}&-\frac{u_{hh}}{2u_{he}^2}&0&0&0&0\\
0&0&0&\frac{1}{u_{hh}}&-\frac{1}{u_{hh}}&-\frac{1}{u_{he}}&0\\
0&0&0&-\frac{1}{u_{hh}}&\frac{1}{u_{hh}}&-\frac{1}{u_{he}}&0\\
0&0&0&-\frac{1}{2u_{he}}&-\frac{1}{2u_{he}}&-\frac{u_{hh}}{2u_{he}^2}&0\\
0&0&0&0&0&0&0\\
\end{array}
\right), \eea
\end{widetext}
and
\begin{widetext}
\bea \label{eq:kernel} \underline{K} (q, \Omega)  = \left(
\begin{array}{ccccccc}
\frac{1}{u_{hh}}+\Pi^{11}_{h_1h_1}&-\frac{1}{u_{hh}}&-\frac{1}{u_{he}}&-\Pi^{12}_{h_1h_1}&0&0&\Pi^{13}_{h_1h_1}\\
-\frac{1}{u_{hh}}&\frac{1}{u_{hh}}+\Pi^{11}_{h_2h_2}&-\frac{1}{u_{he}}&0&-\Pi^{12}_{h_2h_2}&0&\Pi^{13}_{h_2h_2}\\
-\frac{1}{2u_{he}}&-\frac{1}{2u_{he}}&-\frac{u_{hh}}{2u_{he}^2}+\Pi^{11}_{ee}&0&0&-\Pi^{12}_{ee}&\Pi^{13}_{ee}\\
-\Pi^{21}_{h_1h_1}&0&0&\frac{1}{u_{hh}}+\Pi^{22}_{h_1h_1}&-\frac{1}{u_{hh}}&-\frac{1}{u_{he}}&-\Pi^{23}_{h_1h_1}\\
0&-\Pi^{21}_{h_2h_2}&0&-\frac{1}{u_{hh}}&\frac{1}{u_{hh}}+\Pi^{22}_{h_2h_2}&-\frac{1}{u_{he}}&-\Pi^{23}_{h_2h_2}\\
0&0&-\Pi^{21}_{ee}&-\frac{1}{2u_{he}}&-\frac{1}{2u_{he}}&-\frac{u_{hh}}{2u_{he}^2}+\Pi^{22}_{ee}&-\Pi^{23}_{ee}\\
\Pi^{31}_{h_1h_1}&\Pi^{31}_{h_2h_2}&2\Pi^{31}_{ee}&-\Pi^{32}_{h_1h_1}&-\Pi^{32}_{h_2h_2}&-2\Pi^{32}_{ee}&M\\
\end{array}
\right)
\eea
\end{widetext}
Here $M=-\frac{1}{N_0
V_q}+\Pi^{33}_{h_1h_1}+\Pi^{33}_{h_2h_2}+2\Pi^{33}_{ee}$.

The dispersions of seven collective excitations are obtained
from the condition ${\text Det} \underline{K}(q, \Omega) =0$.

To adequately describe the full spectrum of all long-wavelength
collective modes, one has to expand in  $v_F q/\Delta$, but allow
frequency to be of order of $\Delta$ (see Ref. \onlinecite{LB} and
Appendix C).  Our goal, however, is limited: we want to find the plasmon mode in 2D
and the modes which soften at the boundaries of the TRSB state.
All these modes are low-energy modes in the long-wavelength limit,
and to capture them in our approach, it is sufficient to use
double expansion in $v_F q/\Delta$ and in $\Omega/\Delta$.
To get other modes (or resonances) one needs to search for frequencies around $2\Delta$.

In the ++ state, $\phi=0$ in equilibrium, and $\delta^I$ and
$\delta^R$ describe phase and magnitude fluctuations,
respectively. One can easily make sure (see Appendix C) that these
two sets of fluctuations decouple and there are no solutions for amplitude fluctuations
at $\Omega \ll \Delta$.

The three orthogonal phase modes are $~\delta^I_a \equiv
\delta^I_1 - \delta^I_2,~ \delta^I_b \equiv \delta^I_1 +
\delta^I_2 + (2/\gamma) \delta^I_3, ~\delta^I_c \equiv  \delta^I_1
+ \delta^I_2 - (2/\gamma) \delta^I_3$, where $\gamma=2u_{he}L_0$
and $L_0=\frac{u_{hh}+\sqrt{u_{hh}^2+16u_{he}^2}}{8u_{he}^2}$. The
mode $\delta^I_b$ is gapped everywhere in the ++ phase. The mode
$\delta^I_c$  describes fluctuations of the overall phase. This
mode is coupled to fluctuations of the electron density $\delta
\rho$ as \bea\label{eq:phase modes}
-\frac{i\Omega}{\Delta} \delta^I_c-\left(\frac{1}{N_0V_q}+8\right)\delta \rho&=&0\nonumber\\
\frac{v^2_F q^2 - 2 \Omega^2}{4\Delta^2} \delta^I_c +\frac{4 i
\Omega}{\Delta}\delta \rho&=&0 \eea The corresponding dispersion
is a 2D plasmon with $\Omega^2_{pl} = \frac{v_F^2q^2}{2}\left( 8
N_0V_q + 1\right)$. Observe that the plasmon frequency remains the
same as in the normal state~\cite{japa_1}. In general,
$\Omega_{pl}$  in a superconductor scales with the density of
superconducting electrons and is sensitive to
disorder~\cite{dirty}. In our case (clean limit), superconducting
density coincides with the full electron density, hence
$\Omega_{pl}$  does not change between normal and superconducting
states.

The mode $\delta^I_a$ describes antisymmetric phase fluctuations
of the gaps on the two hole pockets. The condensation of this mode
signals the transition to the TRSB state. In the static limit,
this mode totally decouples from density fluctuations. Near
$u_{he} = u^{max}_{he}$ we obtained at $q=0$,
$(\Omega_{\delta^I_a})^2 = (8\sqrt{2}/3) (2\Delta/u^{max}_{he})^2
(u_{he} - u^{max}_{he})$. Not surprisingly, the antisymmetric
phase mode softens at the transition point into the TRSB state
(where  $u_{he} = u^{max}_{he}$). We show the behavior of
$\Omega_{\delta^I_a}$ in Fig.\ref{fig:fig2b}). To properly obtain
the dispersion of this mode, one has to do more involved
calculations as the combinations of $\delta^I_1, \delta^I_2,$ and
$\delta^I_3$, which decouple at a finite $q$, are not the same as
at $q=0$. As a result, the dispersions of Leggett-type modes
generally depend on the Coulomb
interaction~\cite{leggett,LR1,babaev}.

Inside the TRSB state, phase and amplitude fluctuations get mixed
up, as was noticed in Refs.\onlinecite{babaev,stanev}. This is
easily seen form Eq. \ref{eq:kernel} as the off-diagonal
components which connect  the real and imaginary parts of the
order parameter fluctuations, are given by $\Pi^{12}$ which are
proportional to $\sin\frac{\phi}{2}$ (see Eq. \ref{eq:pi_full})
and are non-zero once $\phi \neq 0,\pi$.

The mode which corresponds to the overall phase change is
now $-(\delta^R_1-\delta^R_2)\sin\frac{\phi}{2} +
(\delta^I_1+\delta^I_2)\cos\frac{\phi}{2}
-\frac{2}{\gamma}\delta^I_3$, where in the TRSB state $\gamma = 2
(u_{he}/u_{hh}) \cos \frac{\phi}{2}$, and $\phi$ is given by Eq.
(\ref{ch_1}). This mode decouples from other phase and magnitude
modes, but again couples to $\delta \rho$  and remains a 2D
plasmon.
We solved for the remaining modes and found that the mode
$\delta^I_1-\delta^I_2$, which described antisymmetric phase fluctuations of $\Delta_{h_1}$ and
$\Delta_{h_2}$) outside the TRSB region and softened at the upper boundary of the TRSB state,
 acquires a new functional form inside the TRSB state, and
gets gapped, as expected. As $u_{he}$ decreases and $\phi$
increases and approaches $\pi$, another mode, indicated as the
$M_2$ mode in Fig.\ref{fig:fig2b}, gets soft. This mode is a
coupled oscillation of  $\delta^R_3$ and $\delta_1^R+\delta_2^R$.
The first describes longitudinal fluctuations of the electron gap,
which vanishes at the lower boundary of TRSB state, the second
describes antisymmetric  phase fluctuations of the two hole gaps
(for $\phi = \pi-2 {\tilde \phi}$ and ${\tilde \phi} <<1$,
$\Delta_{h_1}\rightarrow \Delta e^{i(\frac{\pi}{2} -{\tilde
\phi})} \approx \Delta (i + {\tilde \phi})$,  and
$\Delta_{h_2}\rightarrow \Delta e^{-i(\frac{\pi}{2}-{\tilde
\phi})} \approx \Delta (-i + {\tilde \phi})$,  and   $\delta_2^R +
\delta_1^R = 2\delta_2^R= 2{\tilde \phi}$  describes small
deviations from the $+-$ state ). The calculation of this mode
requires some extra care because electron gap $\Delta_e$ vanishes
at the lower boundary of TRSB state, and the expansion in
$\Omega^2/(2\Delta_e)^2$ is only valid if the mode frequency is
below $2\Delta_e$ (Ref.\onlinecite{lara}). Using the formal
expansion in $\Omega$, we obtained the frequency of $M_2$ mode $
\Omega_{M_2} = \sqrt{3} (2\Delta_e)$, which is outside the
applicability limit of the expansion. A more accurate approach is
to keep $\Omega$ along with $\Delta_e$, i.e., replace
$\Omega^2/(2\Delta_e)^2$  by $\Omega^2/(4\Delta^2_e - \Omega^2)$.
This gives $\Omega_{M_2} = (\sqrt{3}/2) (2\Delta_e)$, which is
below the threshold at $2\Delta_e$.  We also found another
low-energy mode using the expansion in $\Omega$, however its
energy is above $2\Delta_e$ even when we keep $\Omega$ along with
$\Delta_e$. This excitation is then inside the continuum and is
not a true collective mode.

We emphasize that the vanishing of $\Delta_e$  is a peculiarity of
the minimal model. In a more general model, the TRSB state emerges
from the modified $+-$ state, in which $\Delta_e$ is already
non-zero. Then it is completely safe to search for soft modes by
expanding in $\Omega/\Delta_i$.

\section{Beyond the minimal model}

 We analyzed whether the TRSB
state survives in more general cases. As a first step, we included
intra-pocket density-density interactions $u_{h_1}$, $u_{h_2}$,
and $u_e$. Applying  the same procedure as before, we found that,
for $u_{h_1}=u_{h_2}$, the phase diagram and the behavior of
collective modes remain the same as in Figs.
\ref{fig:trsb_schematic} and \ref{fig:fig2b},  the only
modification is that at $T=0$ the lower boundary of the TRSB state
now shifts to a finite $u^{min}_{he}=\sqrt{\frac{u_eu_{hh}}{2}}$.
The upper boundary becomes $u_{he}^{max}\approx
u_{he}^0\left[1+\frac{u_{hh}}{4}\frac{(1-\frac{u_{h_1}}{u_{hh}})^2}{\chi^2}ln\left(\frac{2}{\chi}\right)\right]$,
where $\chi=\left( \sqrt{1-\frac{u_{h_1-u_e}}{u_{hh}}} \right)$
and $u_{he}^0 = \frac{u_{hh}}{\sqrt{2}} \chi$ is the point at
which TRSB state emerges right at $T_c$.

When $u_{h_1} \neq u_{h_2}$, the phase diagram  changes
qualitatively (see Fig. \ref{fig:trsb_schematic} b). Now one of the
hole gaps continuously evolves from negative to positive along the $T_c$ line, passing
through zero in between (see Appendix A for details).
 The TRSB state still emerges, but  at
a lower $T$, and survives as long as intra-pocket interactions remain small
compared to $u_{hh}$ (see Appendix B).
To simplify the presentation, we
consider the representative case when $u_{h_2}, u_e=0$ and $u_{h1} \ll u_{hh}$
to understand the changes to the phase diagram. The phase diagram for a generic
$u_{h_1} \neq u_{h_2}$ is qualitatively the same as in the case we considered.

We found that TRSB
state at $T=0$ now exists in an interval between $u^{min}_{he}=0$
and $u^{max}_{he}\approx\frac{u_{hh}}{\sqrt{2}}\left( 1 -
\frac{u_{hh}}{4}ln\left[2/(1+e^{-u_{h_1}/u_{hh}^2})^2\right]\right)$.

\begin{figure}[htp]
$\begin{array}{cc}
\includegraphics[width=1.7in]{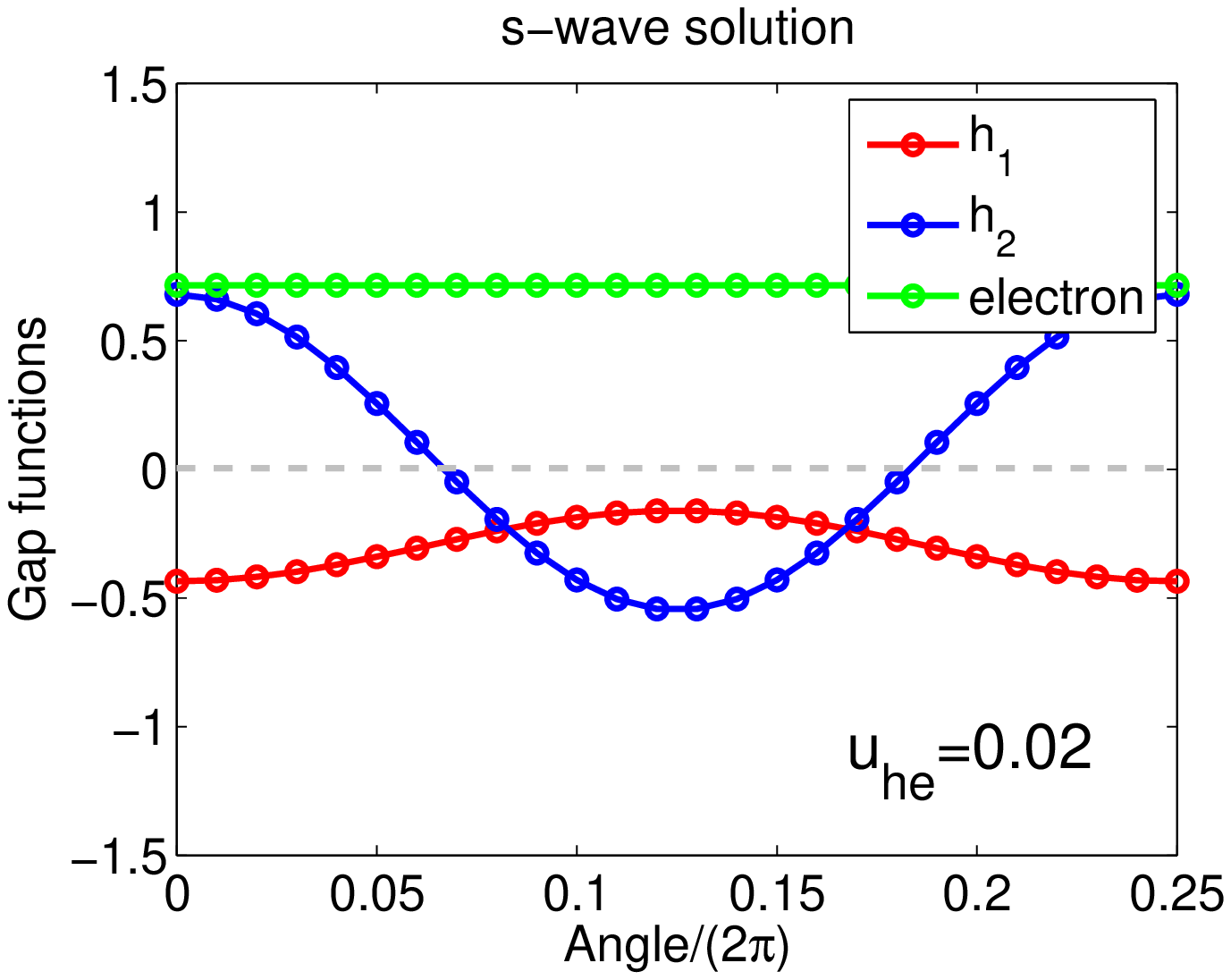}&
\includegraphics[width=1.7in]{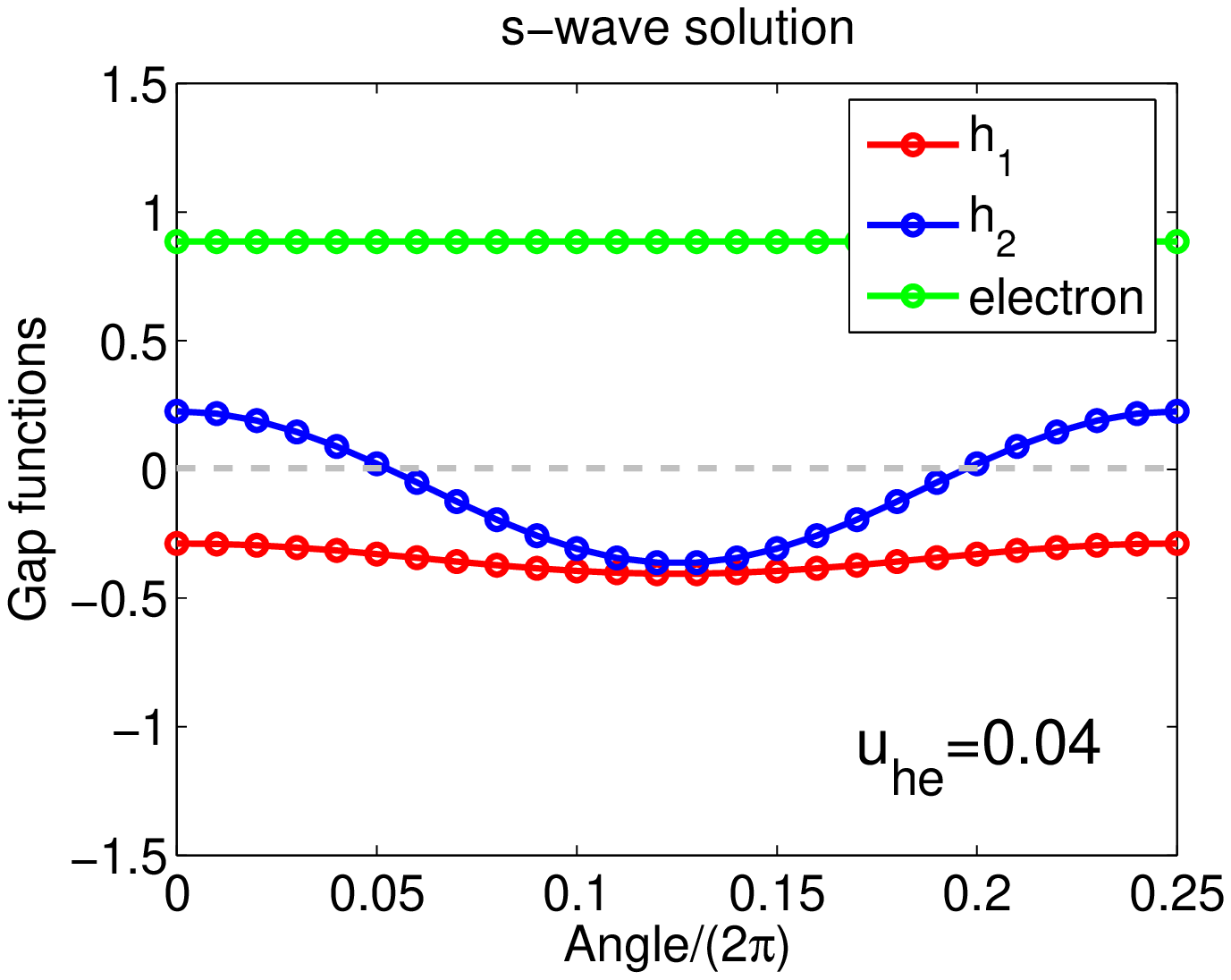}\\
\includegraphics[width=1.7in]{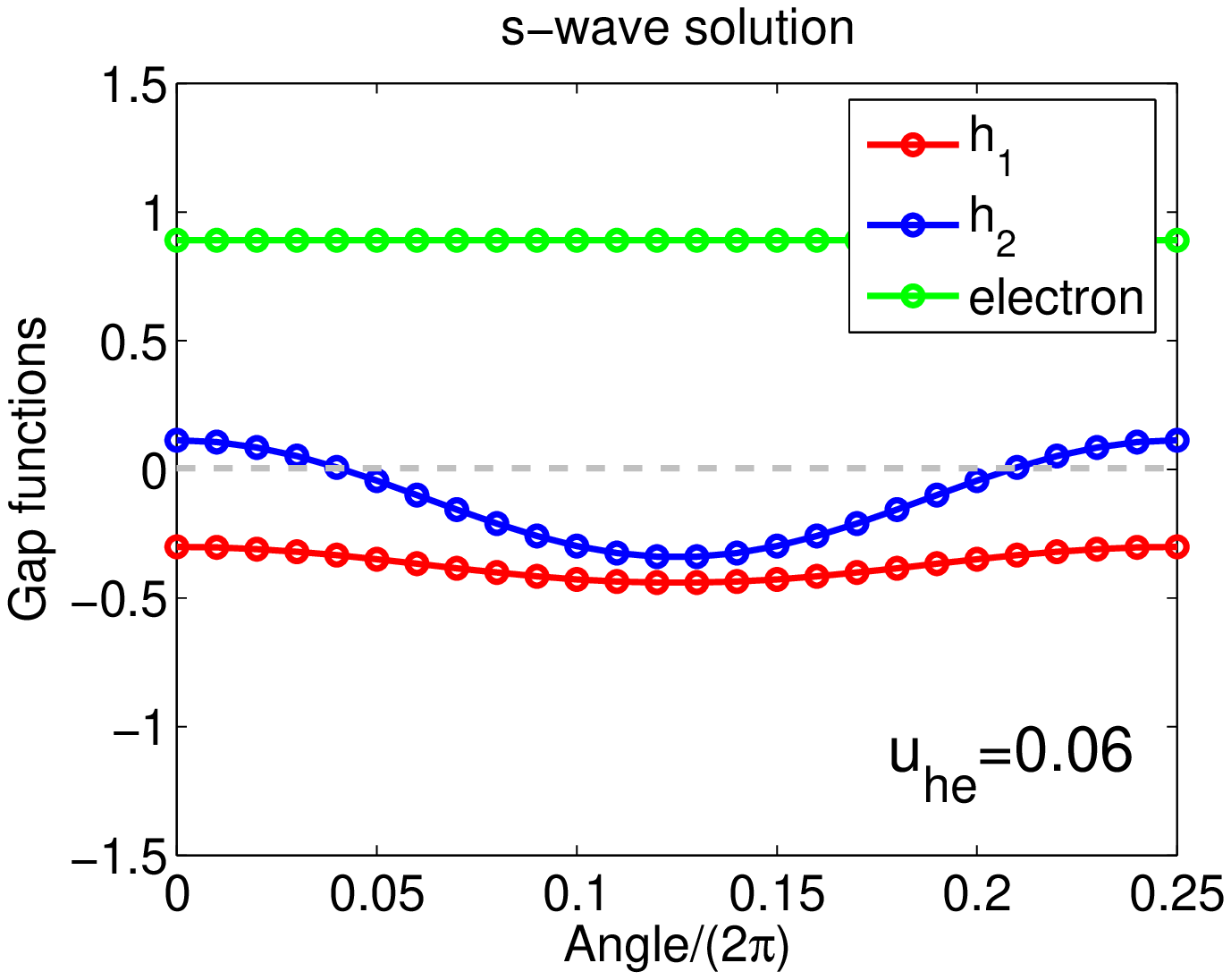}&
\includegraphics[width=1.7in]{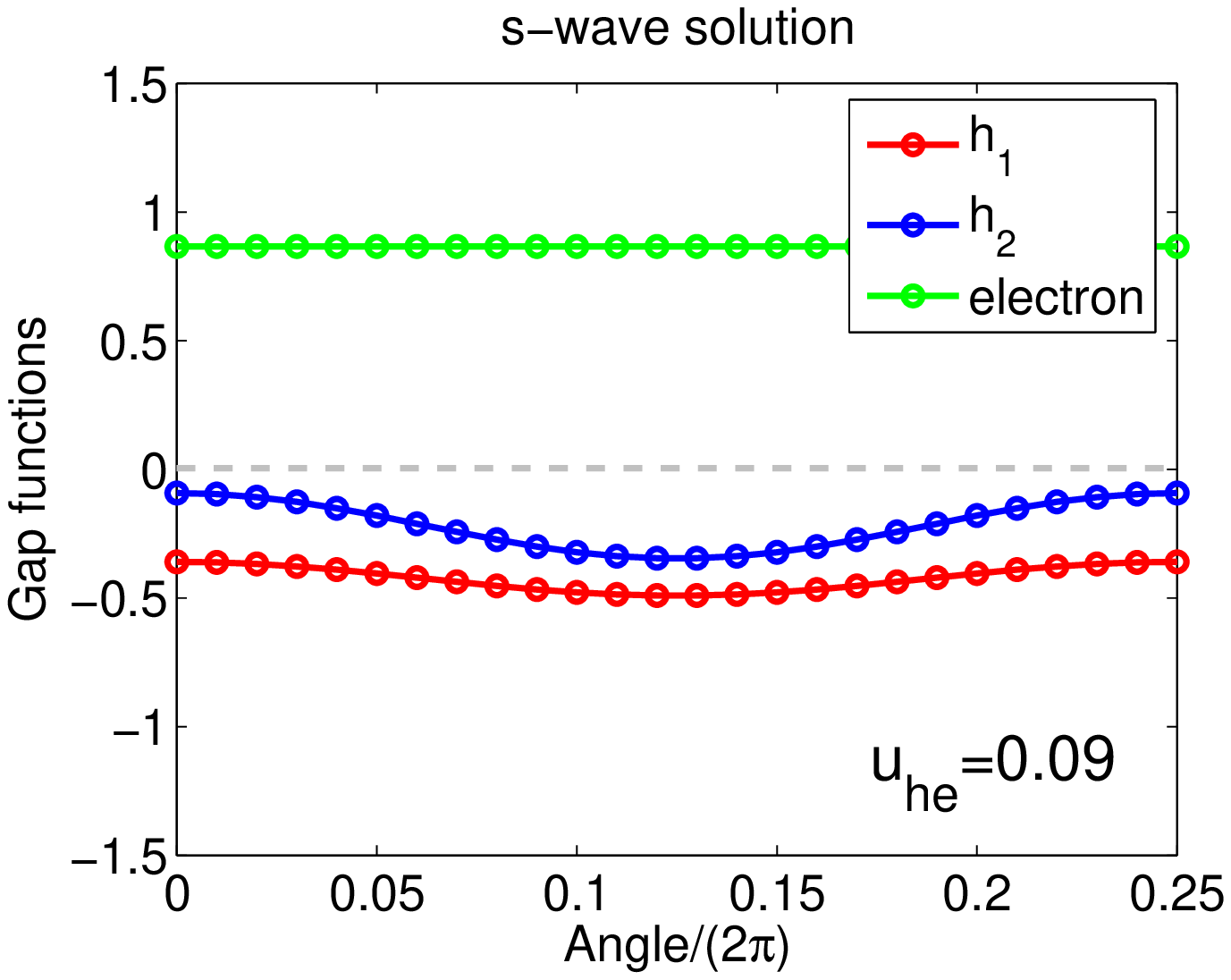}
\end{array}$
\caption{\label{fig1} The $+-$ to ++ transition at $T_c$, with
increasing $u_{he}$ for the case when the two hole pockets are not
equivalent and the interaction $u_{hh}$ has $\cos 4 \theta$
angular dependence. The solutions of the linearized gap equations
are shown for
$u_{he}=0.02$,$u_{he}=0.04$,$u_{he}=0.06$,$u_{he}=0.09$ from left
to right and top to bottom, respectively.  Other parameters are
$\alpha=0.05$,$u_{h_1}=0.2$,$u_{h_2}=0.26$, $u_{hh}=0.2$. Note how
one of the hole gap's average value gets smaller as $u_{he}$
increases, goes through zero, and re-appears with the opposite
sign and with small angular variation at larger $u_{he}$. We
expect such behavior immediately below $T_c$ line in
Ba$_{1-x}$K$_x$Fe$_2$As$_2$, as $x$ decreases from one.}
\end{figure}

We also considered anisotropic  inter-pocket interaction $u_{hh}$
with an extra $\cos 4 \theta$ term, consistent with lattice
symmetry~\cite{LAHA}. This gives rise to $\cos 4 \theta$ angular
variations of $\Delta_{h_1}$ and $\Delta_{h_2}$ and may lead to
accidental gap nodes.  The solution of the set of the gap equations
 for $u_{h1} \neq u_{h2}$ and $u_{hh} (\theta) = u_{hh} (1 + \alpha (\cos{\theta}_{h1} + \cos{\theta}_{h2}))$ is quite involved.
  However, one can show quite generally that  TRSB state is confined to low
temperatures and is separated from the $T_c$ line, like we
previously had for angle-independent interactions.  Immediately
below $T_c$, the $+-$ state gradually evolves  into ++ state,
however,  now only the average value of the``minus" gap goes
through zero at some intermediate $u_{he}$, while the gap itself
does not vanish and just oscillates along the corresponding
pocket.  We illustrate this in Fig \ref{fig1}.

\begin{figure}[htp]
\includegraphics[width=0.9\columnwidth]{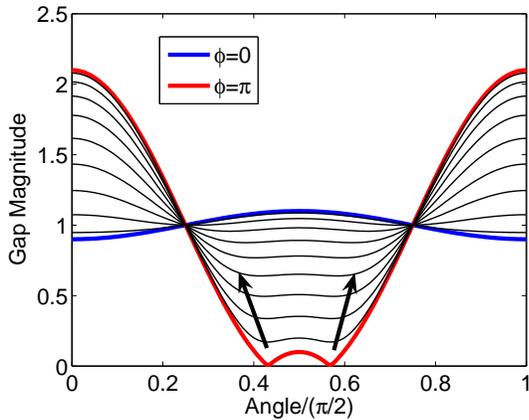}
\caption{\label{fig:fig2c} The gap evolution in the TRSB state for
angle-dependent interactions and two equivalent GCP's, in a
situation when the gap in $+-$ state has accidental nodes and the
gap in $++$ state is nodeless. The nodes disappear once the system
enters the TRSB state, but deep minima (shown by arrows) remain in
some range of $\phi$ ( or equivalently $u_{he}$).}
\end{figure}

Inside the TRSB state at $T < T_c$, the
 number of coupled gap equations equals to nine because in general $\Delta_{h_i} =
\Delta_i\left(e^{i\phi_{ia}}+r_ie^{i\phi_{ib}}\cos4\theta_1\right)
$, $i=1,2$. For $u_{h_1}=u_{h_2}$, we find that $\Delta_{h_1} =
\Delta^*_{h_2} = \Delta e^{i\phi/2}\left(1+(r_a e^{-i\phi} +
r_b)\cos4\theta_1\right)$.  For $\phi =0$ (the ++ state),
accidental nodes exist if $|r_a + r_b| >1$, for $\phi =\pi$ (the
+- state), they exist if $|r_a - r_b| >1$. In the TRSB state,
however, $|\Delta_{h_i}|$ doesn't  crosses zero and can only have
gap minima. We illustrate this behavior in Fig. \ref{fig:fig2c}
for the experimentally relevant case when $+-$ state is nodal and
++ state has a full gap. Observe that  the distance between deep
minima gets larger upon entering the TRSB state. This behavior is
consistent with recent laser ARPES studies of doped
Ba$_{1-x}$K$_x$Fe$_2$As$_2$ (Ref.\onlinecite{shin_last}).

\section{Conclusions}

We considered the evolution of the superconducting gap structure
in strongly hole doped Ba$_{1-x}$K$_x$Fe$_2$As$_2$. Near optimal
doping ($x\sim 0.4$) the pairing symmetry is  $s\pm$, with
different gap sign on hole and electron pockets, but the same sign
of the gap on the hole pockets (a ++ state in our terminology). In
pure KFe$_2$As$_2$ ($x=1$), which has only hole pockets, there are
experimental and theoretical arguments for both $d-$wave and
$s-$wave gap, the latter changes sign between the two GCP's (a
$+-$ state). We assumed s-wave gap symmetry for KFe$_2$As$_2$,
consistent with the laser ARPES data~\cite{ARPES}. The issue we
addressed is how a ++ gap on the GCP's transforms into a $+-$ gap
as $x \to 1$. We found that, for identical GCP's, there is
critical point along $T_c$ line at which the system jumps form
$+-$ to ++ state (see Fig. \ref{fig:trsb_schematic}a). At a lower
$T$, the transformation occurs via an intermediate $s+is$, state
in which the gaps on the two GCP's differ in phase by $\phi$ which
gradually involves from $\phi = \pi$ on one end (the $+-$ state)
to $\phi=0$ on the other end (the ++ state). The system
spontaneously chooses either $\phi$ or $-\phi$ and with this
choice breaks time-reversal symmetry.  We computed the dispersion
of collective excitations and found that two Leggett-type
modes soften at the two ends of the TRSB state. We found that the
TRSB state survives even when the two GCP's are non-identical and
also when the gap on hole pockets is angle-dependent, and even
when  $+-$ and/or  ++ states have accidental gap nodes. In the
former case, near $T_c$ the system gradually evolves from the $+-$
to ++ state, but the TRSB state still emerges at a lower T (see
Fig. \ref{fig:trsb_schematic}b). In the second, the nodes get
lifted once the system enters into a TRSB state (but deep minima
remain). The $s+is$ state is not chiral, but e.g., Kerr effect
measurements still should be able to detect the breaking of
time-reversal symmetry. These measurements are clearly called for.

We acknowledge stimulating conversations with L. Benfatto, R.
Fernandes, I. Eremin, A. Finkelstein, P. Hirschfeld, A. Kamenev,
M. Khodas, A. Levchenko, Y. Matsuda, I. Mazin, R. Prozorov, J.P.
Reid, T. Shibauchi,  V. Stanev, L. Taillefer, R. Thomale, V.
Vakaryuk,  and M. Vavilov. We are particularly thankful to L.
Benfatto for pointing out an error in our original calculation of
the collective modes near the lower boundary of TRSB state. The
research has been supported by DOE DE-FG02-ER46900. SM also
acknowledges support form ICAM-DMR-084415.

\appendix
\section{The Free Energy }

We follow a standard procedure and  introduce bosonic fields
$\Delta_{h_1}, \Delta_{h_2}$, and $\Delta_e$, which describe
fluctuations of the superconducting order parameters on the two
hole and one electron pockets.  We decouple four-fermion
interactions using a Hubbard-Stratonovic(HS) transformation,
integrate over fermions, obtain $Z = \int d \Delta_i
e^{-\mathcal{F}[\Delta_i]}$, and analyze $\mathcal{F}[\Delta_i]$
in the saddle-point approximation. For a model with intra-pocket
and inter-pocket interactions within hole pockets ($u_{h_1},
u_{h_2}$ and $u_{hh}$ terms, respectively) and the interaction
between hole and electron pockets ($u_{he}$ term), we obtained
\bea\label{eq:eff action}
\mathcal{F}[\Delta_i]&=&-\frac{1}{2u_{hh}-u_{h_1}-u_{h_2}}\left[-
2\left(|\Delta_{h_1}|^2+|\Delta_{h_2}|^2\right)+
\right.\nonumber\\
&&+2\left(\Delta_{h_1}\Delta_{h_2}^*+\Delta_{h_1}^*\Delta_{h_2}\right) \nonumber\\
&&+\frac{2(u_{hh}-u_{h_2})}{u_{he}}\left(\Delta_{h_1}\Delta_{e}^*+\Delta_{h_1}^*\Delta_{e}\right)\nonumber\\
&&+\frac{2(u_{hh}-u_{h_1})}{u_{he}}\left(\Delta_{h_2}\Delta_{e}^*+\Delta_{h_2}^*\Delta_{e}\right)\nonumber\\
&&\left. +\frac{2(u_{h_1}u_{h_2}-u_{hh}^2)}{u^2_{he}}|\Delta_{e}|^2\right]\nonumber\\
&&-2L \sum_x |\D_x|^2 +\int G^2\tilde{G}^2 \sum_x |\D_x|^4 \eea
where $L\equiv \int G \tilde {G} \,\,\sim ln\frac{2\Lambda}{T}$,
the sum over $x$ runs over two hole and two electron pockets, and
$G=(i\omega-\e)^{-1}$ and $\tilde{G}=(i\omega+\e)^{-1}$.

Let us first consider the case $u_{h_1}=u_{h_2}=0$. Then one can
diagonalize the quadratic part of the Free energy by introducing
\bea\label{eq:diag_equal_1}
\phi_1&=&\cos\Theta\frac{\Delta_{h_1}+\Delta_{h_2}}{2}-\sin\Theta
\, \Delta_e\nonumber\\
\phi_2&=&\sin\Theta\frac{\Delta_{h_1}+\Delta_{h_2}}{2}+\cos\Theta
\, \Delta_e\nonumber\\
\phi_3 &=&\frac{\Delta_{h_1}-\Delta_{h_2}}{2} \eea where
$\cos\Theta=1/\sqrt{1+\zeta^2}$,
$\sin\Theta=\zeta/\sqrt{1+\zeta^2}$,
 and $\zeta=\frac{u_{hh}}{4u_{he}}\left(1+\sqrt{1+\frac{16
u^2_{he}}{u^2_{hh}}}\right)$.  The action in terms of $\phi_i$
takes the form \bea\label{eq:diag_equal}
\Delta\mathcal{F}_{(2)}[\phi_i]&=&\lambda_1|\phi_1|^2+\lambda_2
|\phi_2|^2 + \lambda_3 |\phi_3|^2, \eea where
\bea\label{eq:diag_equal_2}
\lambda_1&=&\frac{u_{hh}}{2u_{he}^2}\left(1+\sqrt{1+\frac{16 u^2_{he}}{u^2_{hh}}}\right)-4L\nonumber\\
\lambda_2&=&\frac{u_{hh}}{2u_{he}^2}\left(1-\sqrt{1+\frac{16 u^2_{he}}{u^2_{hh}}}\right)-4L\nonumber\\
\lambda_3&=&\frac{4}{u_{hh}}-4L\nonumber\\
\eea

Since $\lambda_2$ is strongly negative, the HS transformation for
$\phi_2$ does not make sense. Because this field does not condense
on physics grounds, we just set $\phi_2 =0$ (see Ref.
\onlinecite{LB2} for more discussion on this). The two other
$\lambda$'s change sign at some, generally different,
temperatures, which depend on $u_{he}/u_{hh}$. When this happens,
either $\phi_1$ or $\phi_3$ condense, depending on whether
$\lambda_1$ or $\lambda_3$ changes sign first upon lowering $T$,
i.e., increasing $L$. (This procedure is formally equivalent to
diagonalizing the linearized gap equation to identify the state
with the leading eigenvalue which in this case would correspond to
either the field $\phi_1$ or $\phi_3$.) The condensation of
$\phi_1$, with $\phi_2 = \phi_3 =0$  brings the system into a ++
phase ($\Delta_{h_1} = \Delta_{h2} = -\Delta_e/\gamma$), while the
condensation of $\phi_3$ with $\phi_2=\phi_1=0$ brings the system
into a $+-$ phase ($\Delta_{h_1} = - \Delta_{h_2}$, $\Delta_e
=0$). At $u_{he}=u_{hh}/\sqrt{2}$, $\lambda_1$ and $\lambda_2$
reach zero at the same $T$, and $\phi_1$ an $\phi_3$ condense
simultaneously (for this $u_{he}$, $\cos\Theta=1/\sqrt{3}$). The
relative magnitude and the relative phase between $\phi_1$ and
$\phi_3$ are decided by minimizing the quartic terms in the Free
energy. Plugging in $\Delta_i$ in terms of $\phi_i$ into Eq.
\ref{eq:eff action}, neglecting $\phi_2$, and using
$u_{he}=u_{hh}/\sqrt{2}$ we obtain \bea\label{eq:final form}
\Delta\mathcal{F}_{(4)} [\phi_i]& = & K_1\left(
|\phi_1|^2+|\phi_3|^2 \right)^2+
K_2\left|\phi_1^2+\phi_3^2\right|^2 + K_3|\phi_1|^4\nonumber\\\eea
where $K_1=\frac{C}{3}$, $K_2=\frac{C}{6}$, $K_3=-\frac{2C}{9}$,
and $C
>0$. The  $K_1$ term is isotropic, the $K_3$ term depends on the
relative magnitudes of $\phi_1$ and $\phi_3$ fields, and the $K_2$
term $K_2|\phi^2_1 + \phi^2_3|^2 = K_2 \left(|\phi_1|^4 +
|\phi_3|^4 + 2 |\phi_1|^2|\phi_3|^2 \cos 2 \theta\right)$. depends
on the relative magnitude and the relative phase $\theta$ between
$\phi_1$ and $\phi_3$: A positive $K_2$ (our case) selects $\theta
= \pm \pi/2$, i.e., if one condensate is real, another is purely
imaginary. Solving for the amplitudes we find $|\phi_3|^2 =
|\phi_1|^2 (1 + K_3/2K_2) = |\phi_1|^2/3$. The state in which both
$\phi_1$ and $\phi_3$ are present, and the relative phase is not
$0$ or $\pi$ is  our TRSB state. Eq. \ref{eq:final form} is
presented in the main text with $\phi_1\rightarrow a$ and
$\phi_3\rightarrow b$.

Away from the degeneracy point the quadratic part of the free
energy takes the form \bea\label{eq:away} \mathcal{F}_{(2)}
[\phi_i]&=&\left(\lambda +
\frac{16}{3}\frac{x}{u_{hh}}\right)|\phi_1|^2 + \lambda |\phi_3|^2
\eea where $\lambda = 4(1/u_{hh}-L)$ and $x = 1 -
\sqrt{2}u_{he}/u_{hh}$. The leading instability to the left of the
degeneracy point (at $x >0$) is into the $\phi_3$ state, and to
the right of it (at $x <0$) it is into the $\phi_1$ state. Once
one order sets in, it acts against the appearance of the other.
Still, we found that, e.g., at $x
>0$, $\phi_1$ still condenses at $\lambda_{cr} = -(16 x/3u_{hh})
((K_1 + K_2)/2K_2) = -(16 x/3u_{hh}) *(3/2)$. The corresponding
temperature $T_{cr}$ is smaller than without $K$ terms, but is
still finite. Once $\phi_1$ becomes non-zero, a positive $K_2$
again selects a relative phase of $\pm \pi/2$ between $\phi_3$ and
$\phi_1$ (which corresponds to the $\phi=\pi$ boundary for the
TRSB state). This consideration leads to the phase diagram in Fig.
1a in the main text.

We extended this analysis to the case when $u_{h_1}=u_{h_2}\neq 0$
and found the same results as above. However, when $u_{h_1}\neq
u_{h_2}$, the phase diagram changes qualitatively. To show the new
physics and at the same time avoid lengthy formulas,  we set
$u_{h_1} \ll u_{hh};\, u_{h_2}=0$ and consider $u_{he}$ near
$u_{hh}/\sqrt{2}$, at which ++ and $+-$ phases cross  at $T_c$.
Specifically, we set  $u_{h_1} =2 y u_{hh}$, $u^2_{he} =
(u^2_{hh}/2) (1 + 2 c y)$, and obtained the phase diagram to first
order in $y \ll 1$.

At a non-zero $y$, the quadratic part of the Free energy reads
\bea\label{eq:enuqual} \mathcal{F}_2 [\phi_i] &=& 4 \left(\frac{1+y(1-4c)/3}{u_{hh}} -L\right)|\phi_1|^2 \nonumber \\
&& + 4
\left(\frac{1+y}{u_{hh}}
-L\right)|\phi_3|^2 \nonumber \\
&&-  4 \left(\frac{1+y(7-10c)/3}{2u_{hh}} +L\right)|\phi_2|^2 \nonumber \\
&&- \frac{2 \sqrt{2}y}{\sqrt{3} u_{hh}} \left(\phi^*_3 (\phi_2 -
\sqrt{2} \phi_1) + c.c\right)
\eea
The $\phi_2$ mode is again non-critical, and $\phi_2$ can be sent
to zero. For the remaining two modes, we have

\bea\label{eq:enuqual_1} \mathcal{F}_2[\phi_i] &=&4
\left(\frac{(1+y)}{u_{hh}} -L\right)|\phi_3|^2 \nonumber\\
&+& 4\left(\frac{1+ y(1-4c)/3}{u_{hh}}
-L\right)|\phi_1|^2\nonumber\\
&+&\frac{4y}{\sqrt{3}u_{hh}}\left(\phi_3\phi^*_1 + +c.c\right)
\eea Diagonalizing this quadratic form by \bea \phi_1 = \psi_1
\cos{\eta}  + \psi_3 \sin{\eta}   ~~\phi_3 = - \psi_1 \sin{\eta} +
\psi_3 \cos{\eta}, \label{ch_1_1} \eea we obtain $\tan{2\eta} =
\sqrt{3}/(1+2c)$. Taking the positive root $\tan \eta =
\frac{1}{\sqrt{3}} \left[\sqrt{(1+2c)^2 + 3}-(1+2c)\right]$, we
obtain \bea \label{eq:enuqual_2}
 &&\mathcal{F}_2[\psi_i]= \nonumber \\
 && 4
\left[\left(\frac{1 +
\frac{y}{3}\left(2(1-c)-\sqrt{(1+2c)^2+3}\right)}{u_{hh}}-L\right)|\psi_1|^2\right.
\nonumber\\
&+& \left. \left(\frac{1 +
\frac{y}{3}\left(2(1-c)+\sqrt{(1+2c)^2+3}\right)}{u_{hh}}-L\right)|\psi_3|^2\right]\nonumber\\
\eea

\begin{figure}[htp]
$\begin{array}{cc}
\includegraphics[width=0.5\columnwidth]{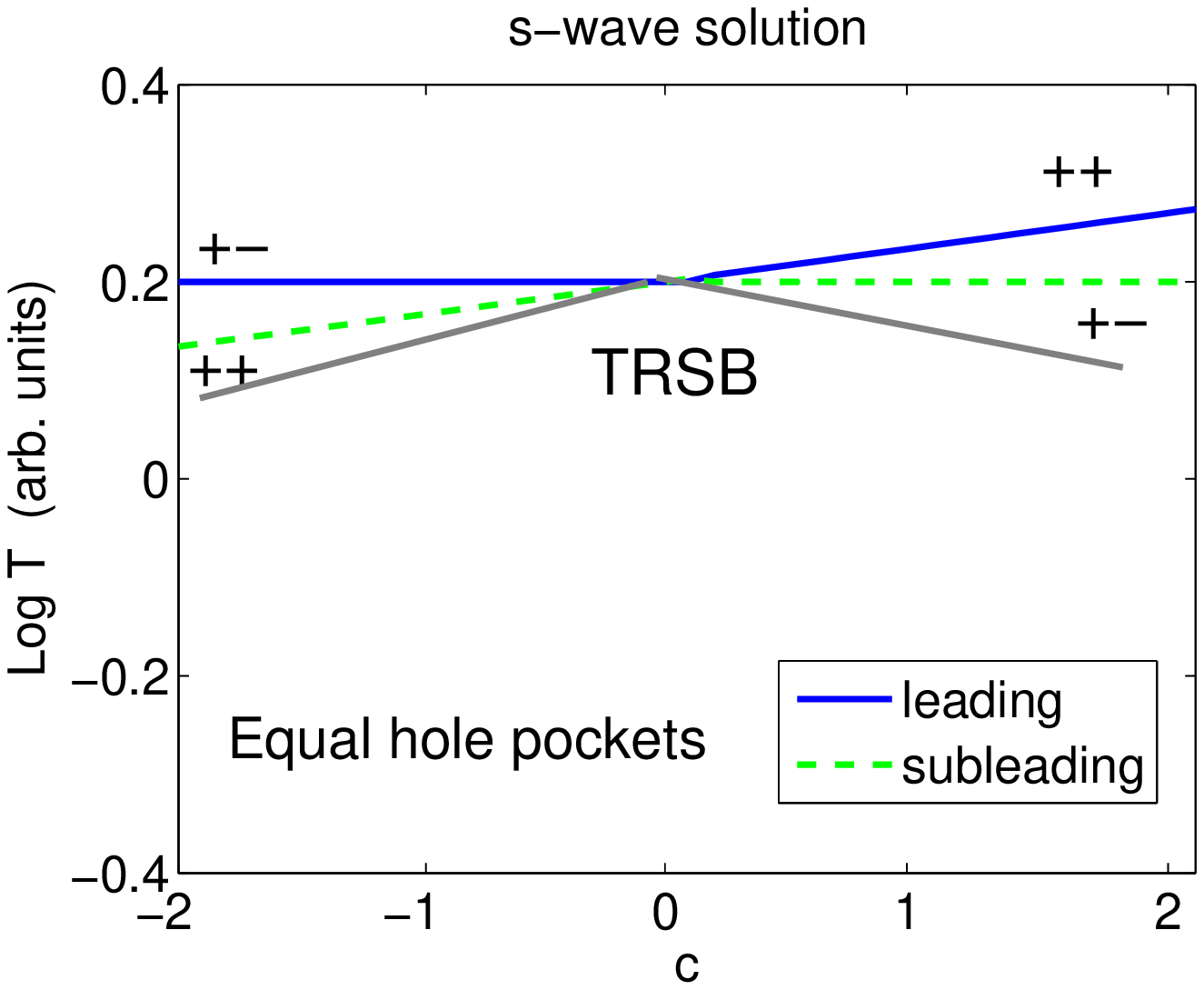}&
\includegraphics[width=0.5\columnwidth]{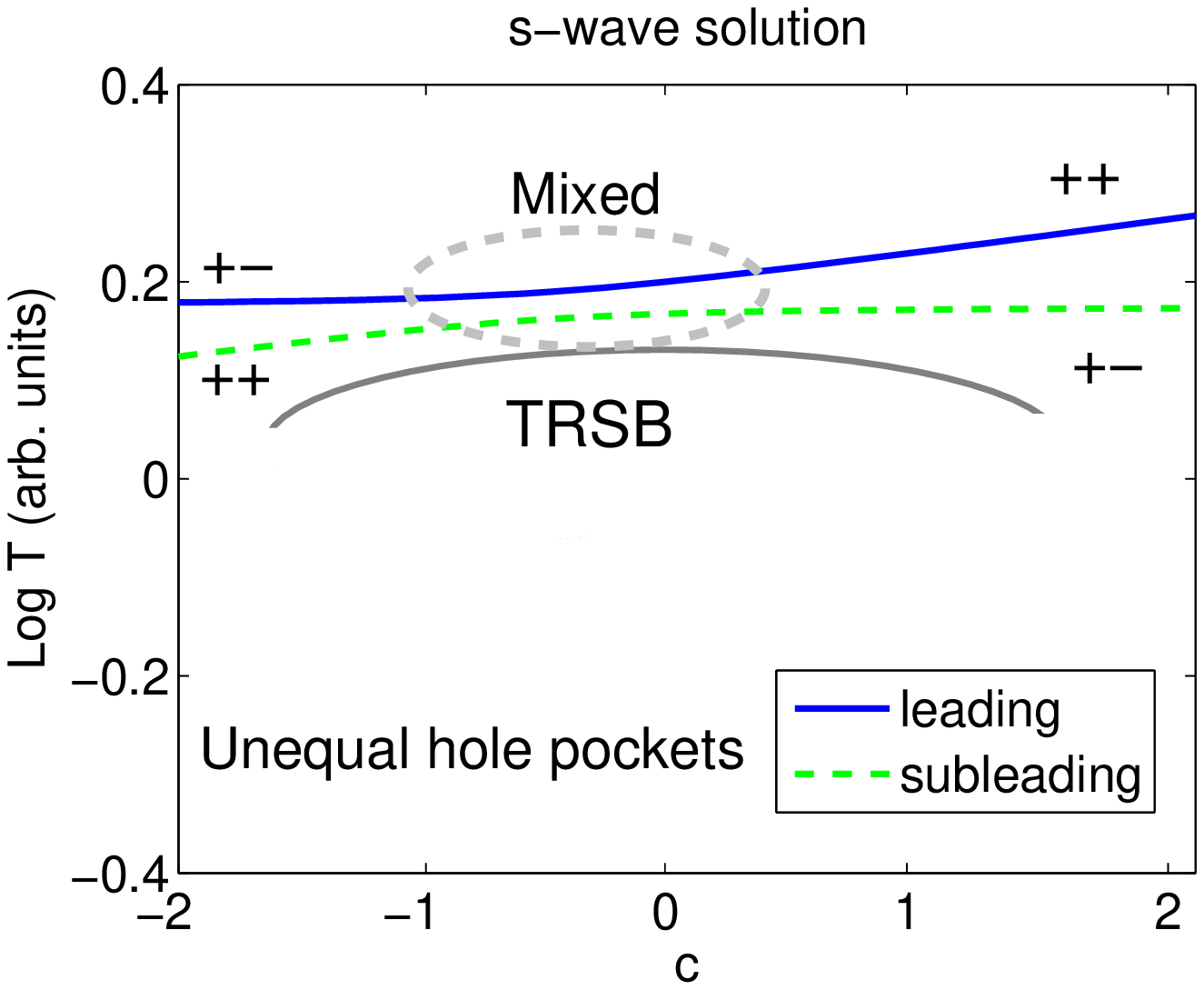}
\end{array}$
\caption{\label{fig:EV} Temperatures (T) at which the prefactors
for the quadratic terms in Ginzburg-Landau expansion for the two
critical fields change sign. The parameter $c$ measures the
deviation of hole-electron interaction $u_{he}$ from the critical
value ($=u_{hh}/\sqrt{2}$).  Left panel -- two equivalent hole
pockets ($y=0$). In this situation, the condensation of one
critical field leads to $+-$ order, the condensation of the other
leads to ++ order.  The two lines cross at the critical $u_{he}$.
In the presence of mode-mode coupling, the emergence of one order
tends to prevent the emergence of the other, and the actual
temperature, at which the second order emerges, gets smaller
(black line).  We found (see text) that below the black line the
two orders lock into TRSB state.  Right panel - non-equivalent
hole pockets ($y=1/8$). The eigenfunctions reduce to pure $+-$ and
++  only at large $|c|$, while in the region labeled `mixed' the
system gradually transforms form the $+-$ to ++ order with one of
the hole gaps going through zero in between.  The lines at which
the prefactors for the quadratic terms vanish now do not cross.
Due to mode-mode coupling, the order which appears first now
induces second order, i.e., both are present immediately below the
actual $T_c$ line, with a relative phase of $0$ or $\pi$, i.e.,
time-reversal symmetry is not broken at $T_c$. The TRSB state
still emerges, but at a lower $T$ (below the black line).}
\end{figure}

We see that the temperatures at which $\psi_1$ and $\psi_3$ modes
condense are different and $\psi_1$ mode condenses first for all
values of $c$. The $\psi_1$ mode condenses at $L_{\psi_1} = (1+
S_1(c))/u_{hh}$, where $S_1(c) = (8/3) (1-c -\sqrt{1 + c +c^2})$,
and the $\psi_3$ mode condenses at $L_{\psi_3} = (1+
S_3(c))/u_{hh}$, where $S_3(c) = (8/3) (1-c +\sqrt{1 + c +c^2})$.
We plot the temperatures at which the prefactors for $|\psi_1|^2$
and $|\psi_3|^2$ terms vanish in Fig \ref{fig:EV} The condensation
of $\psi_1$ field leads to a superconducting state in which all
three gaps $\Delta_{h_1}$, $\Delta_{h_2}$, and $\Delta_e$ are
generally present and are different from each other. At large
positive $c$ (i.e., at larger $u_{he}$) the state immediately
below the condensation temperature of $\psi_1$ is close to the ++
state, with $\Delta_{h_1} \approx \Delta_{h_2}$ and $\Delta_e$ of
opposite sign compared to $\Delta_{h_1}$ and $\Delta_{h_2}$. At
large negative $c$ (smaller $u_{he}$) the state immediately below
the condensation temperature of $\psi_1$ is close to the $+-$
state, with $\Delta_{h_1} \approx - \Delta_{h_2}$ and smaller
$\Delta_e$. In between, the condensed state is a mixture of ++ and
$+-$ states. In particular, for $c=0$, $\Delta_e
=-\Delta_{h_2}/\sqrt{2}$ and $\Delta_{h_1} =0$, i.e., the gap on
the hole pocket, for which we kept intra-pocket repulsion,
vanishes.  We analyzed the form of the condensate for various $c$
(i.e., various $u_{he}/u_{hh}$) and found a continuous evolution,
in the process of which one of hole gaps gets smaller, passes
through zero, and then re-emerges with the opposite sign.
Specifically, we found, right below $T_c$ for the $\psi_1$ mode,
\bea
&&\Delta_e = - \frac{\Delta_{h1} + \Delta_{h2}}{\sqrt{2}}, \nonumber \\
&&\frac{\Delta_{h1} - \Delta_{h2}}{\Delta_{h1} + \Delta_{h2}} =
(1+2c)-\sqrt{3 + (1+2c)^2} \eea

Without quartic terms, the modes $\psi_1$ and $\psi_3$ are
decoupled and the system undergoes two superconducting transitions
at $L_{\psi_1}$ and $L_{\psi_3}$.  The mode which condenses at
$L_{\psi_3}$ is almost ++ state at large negative $c$, almost $+-$
state at large positive $c$, and a mixed state in between. E.g.,
at $c=0$, the $\psi_3$ condensate has components $\Delta_{h_1}
=-2\Delta_{h_2}, \Delta_e =\Delta_{h_2}/\sqrt{2}$.

The situation changes when we include quartic terms into
consideration.  We use Eq. (\ref{eq:final form}) as an input,
substitute $\phi_{1,3}$ in terms of $\psi_{1,3}$ via (\ref{ch_1}),
and obtain  $\mathcal{F}_4[\psi_i]$. Carrying out the
calculations, we find that the four-fold term contains a linear
piece in $\psi_3$ in the form $2 K_3 \sin{2\eta} \cos^2{\eta}
|\psi_1|^3|\psi_3| \cos\theta_{13}$, where $\theta_{13}$ is a
relative phase between the condensates of $\psi_1$ and $\psi_3$.
This term acts as an "external field" for $\psi_3$ and makes
$\psi_3$ non-zero once $\psi_1$ condenses.  Because $K_3<0$, the
system initially selects $\theta_{13} =0$, i.e., $\phi_3$ field
emerges with the same phase as $\phi_1$.  This implies that the
state immediately below $L_{\psi_1}$ breaks a  U(1) gauge symmetry
(the overall phase gets fixed), but time-reversal symmetry remains
unbroken.  The situation changes, however, when the temperature
gets lower and $\psi_3$ grows. The full dependence of
$\mathcal{F}_4[\psi_i]$ on $\theta_{13}$ is in the form
\bea
&& \mathcal{F}_4[\psi_i] = 2K_3 \cos\theta_{13} \sin{2\eta}  \nonumber\\
&& \times|\psi_1||\psi_3| \left(|\psi_1|^2 \cos^2{\eta} + |\psi_3|^2 \sin^2{\eta}\right) \nonumber \\
&& + \cos^2 \theta_{13} |\psi_1|^2|\psi_3|^2\left(4K_2 + K_3
\sin^2{2\eta}\right) \eea

Analyzing this form, we immediately find that the prefactor for
$\cos^2{\theta_{13}}$ is necessary positive. Minimizing with
respect to $\theta_{13}$, we then find that, at some finite
$\psi_3$, the equilibrium value of $\theta_{13}$ shifts from
$\phi_{13}=0$ to a finite  $\theta_{13} = \pm b$, $b \neq 0$. For
large and small $c$, this happens already at small $\psi_3$, which
are well within the applicability of the expansion in powers of
$\psi$. Thus, for large positive $c$, the critical $|\psi_3|
=|\psi_1|/(\sqrt{3} |1+2c|)$.

Once the system selects a non-zero $\theta_{13}$, it breaks
additional $Z_2$ symmetry by selecting either positive or negative
value of the relative phase $\theta_{13}$. The $Z_2$ breaking then
implies that time-reversal symmetry is broken, i.e., once
$\theta_{13}$ becomes non-zero, the system enters into a TRSB
phase. The region of this phase shrinks as $u_{h_1}$ increases but
definitely remains finite as long as $u_{h_1} << u_{hh}$, i.e., as
long as our parameter $y$ is small.

When both $u_{h_1}$ and $u_{h_2}$ are non-zero, the calculations
become more involved, but the physics remains the same.  We also
analyzed the effect of adding intra-pocket interaction $u_e$ for
electron pockets.  Like in the case of $u_{h_1} =u_{h_2}$, a
non-zero $u_e$ shifts the lower boundary of the TRSB state to a
finite $u_{he}$. There is one new effect compared to the case
$u_{h_1} =u_{h_2}$: because now (when $u_{h_1} \neq u_{h_2}$),
$\Delta_e$ remains non-zero to the left of the lower boundary of
the TRSB state, the mode which describes longitudinal fluctuations
of $\Delta_e$, no longer strongly couples  to antisymmetric phase
fluctuations of the two hole gaps, and the mode which softens at
the lower boundary of TRSB state becomes a pure Leggett-type phase
mode.

\section{Non-linear gap equations at $T=0$}

The key goal of the analysis is to show that TRSB state, which
starts as a point along $T_c$ line, extends to a finite range of
system parameters at $T=0$

\begin{figure}[htp]
\includegraphics[width=0.9\columnwidth]{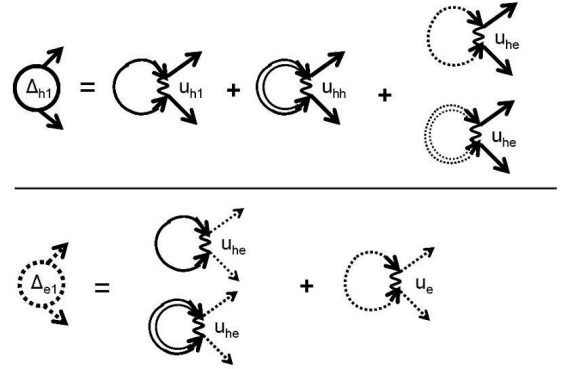}
\caption{\label{fig:gap eq} Diagrammatic representation of the set
of non-linear
equations for the gaps  $\Delta_{h_1}$ and
$\Delta_{e_1}$ (viewed as anomalous self energies). In our case
$\Delta_{e_1}=\Delta_{e_2}\equiv\Delta_e$. The equation for $\Delta_{h_2}$ is similar and  not
shown. The double headed arrows correspond to the anomalous Greens
functions; The single and double solid lines and the single and
double dotted lines are anomalous Green functions  for fermions near the hole
 pockets ($h_{1,2}$) and near  electron  pockets ($e_{1,2}$), respectively.}
\end{figure}

The set of non-linear gap equations in a generic model with
inter-pocket interactions $u_{hh}$, $u_{he}$, and intra-pocket
interactions $u_{h1}$, $u_{h2}$, and $u_e$ is shown
diagrammatically in Fig \ref{fig:gap eq}. Each anomalous vertex is
a gap $\Delta_x$, which, in general, is a complex variable ($x =
h_1, h_2$, and $e$), and each fermionic bubble is a sum of normal
and anomalous Green functions \be G^{(x)}_{\alpha\beta} =-
\delta_{\alpha,\beta} \frac{i\omega+\e_x}{\omega^2+E^2_x}, ~~
 F^{(x)}_{\alpha\beta} = g_{\alpha,\beta}\frac{\Delta_x}{\omega^2+E^2_x}
\ee where $E_x = \sqrt{\epsilon^2_x + |\Delta|^2_x}$, $\epsilon_x$
is the fermionic dispersion near the pocket x, and
$g_{\alpha,\beta} = i\sigma^{y}_{\alpha\beta}$. Evaluating the
diagrams, we obtain at $T=0$ \bea \label{eq:nonlingap eq}
\D_{h_1}&=&-u_{h_1} \D_{h_1}L_1 -u_{hh}\D_{h_2}L_2-2u_{he}\D_e L_e\nonumber\\
\D_{h_2}&=&-u_{hh} \D_{h_1}L_1 -u_{h_2}\D_{h_2}L_2-2u_{he}\D_e
L_e\nonumber\\
\D_e&=& -u_{he}\D_{h_1}L_1-u_{he}\D_{h_2}L_2-u_e\D_e L_e \eea
where $L_x\equiv ln \left( \frac{2\Lambda}{|\D_x|}\right)$

\subsection{The symmetric case}

Consider first the symmetric case $u_{h_1} = u_{h_2}$.  Then
$\Delta_{h_1}=\Delta_{h_2}=\Delta$ and $L_1=L_2=L$. Without loss
of generality, the overall phase can be set such that $\Delta_e$
is real. The two hole gaps must then satisfy $\Delta_{h_2} =
\Delta_{h_1}^*$, i.e in general
$\Delta_{h_1}=\Delta\,e^{i\phi/2}$,
$\Delta_{h_2}=\Delta\,e^{-i\phi/2}$. The electron gap $\Delta_e$
also scales with $\Delta$, and we write $\Delta_e = -\gamma
\Delta$, in which case $L_e\equiv L-ln\gamma$. The three variables
$\Delta, \gamma$, and $\phi$ are the solutions of the set of three
non-linear gap equations (we recall that $L =
\log\frac{2\Lambda}{\Delta}$). We have, from (Eq.
\ref{eq:nonlingap eq}), \bea \label{eq:NLgap eq var1} \left[
1-\left(u_{hh}-u_{h_1}
\right)L\right]\sin(\phi/2) &=&0\nonumber\\
\left[ 1+\left(u_{hh}+u_{h_1}\right)L\right]\cos(\phi/2)&=&
2u_{he}\gamma
L_e\nonumber\\
\left[ 1+u_e L_e\right]\gamma&=&2 u_{he}\cos(\phi/2) L \nonumber\\
\eea For the $+-$ state, $\phi =\pi$, and  we have $\gamma =0$ and
$L = 1/(u_{hh}-u_{h_1})$. For the ++ state, $\phi=0$, $L$ is
approximately the smallest positive solution of
\beq\label{eq:phi=0(b)} 1+(u_e+u_{hh}+u_{h_1}))L+\left(
u_e(u_{hh}+u_{h_1})-4u_{he}^2\right)L^2=0\eeq and $\gamma$ is the
solution of $\gamma (1 + u_e (L - log{\gamma}) = 2 u_{he} L$.

For the TRSB state, $\phi$ is different from $0$ and $\pi$, and we
have \bea\label{eq:sol}
L&=&\frac{1}{u_{hh}-u_{h_1}}\nonumber\\
L_e&=& \frac{u_{hh}}{2u_{he}^2-u_e u_{hh}} \nonumber\\
\gamma &=&2 \frac{u_{he}L}{1+u_e L_e}\cos\frac{\phi}{2} \eea

The upper and lower boundaries of the TRSB state are obtained by
matching the TRSB solution and the solutions for the ++ and $+-$
states, respectively. This gives $u^{max}_{he}$ and
$u^{min}_{he}$, which we presented in the main text.

\subsection{Non equivalent hole pockets}

For  $u_{h_1} \neq u_{h_2}$, $\Delta_{h_i}=\Delta_i
e^{i\phi_i/2}$, and both $\Delta_{1,2}$ and $\phi_{1,2}$ are
generally different. The analysis now involves five variables (two
complex $\Delta_{h_i}$ an one real $\Delta_e$, and is quite
involved. However, less efforts are needed to just prove that TRSB
state exists because near its upper and lower boundaries $\phi_1$
and $\phi_2$ approach zero or differ by $\pi$, respectively,
and one can expand in the deviations from equilibrium
$\phi_i$'s.

As an example, consider the system near the upper boundary of the
TRSB state. Here $\phi_1$ and $\phi_2$ are both small. Expanding
in the set of complex equations (\ref{eq:nonlingap eq})  for
$\Delta_{h_i}$ and $\Delta_e$ to linear order in $\phi_{1,2}$, and
separating real and imaginary parts, we obtain, from the imaginary
parts, \bea \label{eq:im gap eq1}
 &&\Delta_1 \left(1+u_{h1} L_1\right) \phi_1 + L_2 \Delta_2 \phi_2 =0 \nonumber \\
 &&\Delta_2 \left(1+u_{h2} L_2\right) \phi_2 + L_1 \Delta_1 \phi_1 =0 \nonumber \\
 &&\Delta_1  L_1 \phi_1 + \Delta_2 L_2 \phi_2 =0
 \eea
Combining, e.g., the first two and the last two equations and each
time setting the  determinant to be zero and combining with the
third equation in (\ref{eq:im gap eq1}), we immediately obtain
\bea
L_1&=& \log\frac{2\Lambda}{\Delta_1} = \frac{1}{u_{hh}-u_{h_1}}, \nonumber \\
L_2&=& \log\frac{2\Lambda}{\Delta_2} = \frac{1}{u_{hh}-u_{h_2}},
\label{ch_3} \eea

The real parts of the same set of Eqs.(\ref{eq:nonlingap eq}) can
be evaluated at $\phi_{1} = \phi_2 =0$. The first two equations of
the set (\ref{eq:nonlingap eq}) with real $\Delta_{h_i} =\Delta_i$
are identical for $L_{1,2}$ (and $\Delta_{1,2}$) given by
(\ref{ch_3}) and using them we can express $\Delta_e L_e =
\Delta_e \log\frac{2\Lambda}{|\Delta_e|}$ in terms of various
couplings $u$.  Solving for $\Delta_e$ and substituting the result
into the last equation in (\ref{eq:nonlingap eq}) we obtain the
expression for $u_{he} = u^{max}_{he}$ for the upper boundary of
the TRSB state. The result for $u^{max}_{he}$ for $u_{h_2} = u_e
=0$ and $u_{h_1} << u_{hh}$ is presented in the main text. The
result for the lower boundary of the TRSB state, $u^{min}_{he}$ is
obtained in a similar manner, by expanding near $\phi_{1,2} =
\pi$.

\subsection{TRSB state for angle-dependent interaction}

Our primary interest is to study how the TRSB state is modified if
outside this state the gaps on the two $\Gamma$-centered hole
pockets have angular dependence and even accidental nodes, if this
dependence is strong enough.

To focus on this physics and avoid lengthy formulas, we ignore
potential anisotropy of intra-pocket interactions $u_{h_i}$ and
$u_e$ and of electron-hole interaction $u_{he}$, and only include
the anisotropy of the interaction $u_{hh}$ between the two
$\Gamma$-centered hole pockets. By symmetry~\cite{LAHA},
angle-dependence of $u_{hh}$ comes in the form \beq\label{eq:uh
angle dep} u_{hh}(k,p)=u_{hh}\left( 1+2\alpha\cos4\theta_k +
2\alpha\cos4\theta_p \right) + ... \eeq where dots stand for $\cos
8\theta$, etc terms which we neglect. The most general solution
for the hole gaps for this form of the interaction is
\bea\label{eq:gen gap sol structure}
\Delta_{h_1}&=&\Delta_1\left(e^{i\phi_{1a}}+r_1e^{i\phi_{1b}}\cos4\theta\right)\nonumber\\
\Delta_{h_2}&=&\Delta_2\left(e^{i\phi_{2a}}+r_2e^{i\phi_{2b}}\cos4\theta\right)\nonumber\\
\Delta_e &=&\Delta_3 \eea
where without loss of generality we can set $\Delta_i$ and $r_i$ to be positive.
As before, we select  $\Delta_e$ to be real by adjusting the overall phase.

To obtain the gaps in the TRSB state for arbitrary interactions
$u$, one has to solve the set of nine coupled equations, which can
only be done numerically.  One can, however, still find an
analytical solution for the case $u_{h_1}=u_{h_2}$. In this
situation, two hole pockets are equivalent, and one can easily
show that $\Delta_{h_1}=\Delta^*_{h_2}$. We verified that the set
of non-linear gap equations is satisfied if we use the following
ansatz \bea\label{eq:gen} \Delta_{h_1}& = & \Delta^*_{h_2} =
\Delta e^{i\phi/2}\left(1+(r_a
e^{-i\phi} + r_b)\cos4\theta\right)\nonumber\\
\Delta_e& = &-\gamma\Delta\eea
This ansatz contain five unknowns ($\Delta, \gamma, r_a, r_b, \phi$).
Substituting these forms into the set of non-linear gap equations
(Eq. (\ref{eq:nonlingap eq}) with $u_{hh}$ given by (\ref{eq:uh angle dep}), we obtain
\bea\label{eq:angle}
 r_a&=&-2\alpha u_{hh} \int L_{\theta}(1+r_b
\cos4\theta)\nonumber\\
r_b&=&-2\alpha u_{hh} r_a\int L_{\theta}\cos4\theta \nonumber \\
\cos\frac{\phi}{2}&=&-\int
(u_{hh}A_{\theta}+u_{h_1})L_{\theta}(1+r_b
\cos4\theta)\cos\frac{\phi}{2} \nonumber\\
&-&r_a\int (u_{hh}A_{\theta}+u_{h_1})L_{\theta}
\cos4\theta\cos\frac{\phi}{2} \nonumber\\
&+&2u_{he}L_e\gamma\nonumber\\
1&=&\int (u_{hh}A_{\theta}-u_{h_1})L_{\theta}(1+r_b
\cos4\theta) \nonumber\\
&-&r_a\int (u_{hh}A_{\theta}-u_{h_1})L_{\theta} \cos4\theta \nonumber \\
\gamma&=&2u_{he}\int L_\theta\cos\frac{\phi}{2}\left(1 + (r_a +
r_b)\cos4\theta\right) \eea where
$L_{\theta}=ln\frac{2\Lambda}{|\Delta(\theta)|}$ and
$A_{\theta}=1+2\alpha\cos4\theta$. When $\alpha=0$, we have
$r_a=r_b=0$, and the other three equations coincide  with what we
had in the isotropic case.

We analyze this set both analytically and numerically, and found
that TRSB state (the one with $\phi$ different from zero or $\pi$)
still exists, at $T=0$,  in some range of $u_{he}$, even if the
hole gaps in $+-$ and/or ++ states  have accidental nodes.
However, in the TRSB state, the gap amplitude has minima but no
nodes, simply because $|\Delta_{h_1}| = |\Delta_{h_2}| = \Delta^2
((1 + (r_a \cos{\phi} + r_b)\cos{4 \theta})^2 + r^2_a \sin^2{\phi}
\cos^2 {4\theta})$ never hits zero when $\sin{\phi}$ is non-zero.
We discuss this in the main text.

\section{Collective modes}

In this Appendix we  present some details of the derivation of the
dispersion of collective modes.

We consider the minimal model with two equal hole pockets and two
inter-pocket interactions $u_{he}$ and $u_{hh}$. The extension to
more general cases is straightforward, but the formulas become
more cumbersome.

We include both the pairing interactions ($u_{he}$ and $u_{hh}$)
and 2D long-range Coulomb interaction $V_q = A_2/|q|$, $A_2 = 2\pi
e^2$. To obtain the dispersion of collective modes, we add to the
system a small frequency and momentum-dependent perturbation (the
bare terms) \bea\label{eq:pert} H_{pert} &=&\sum_k \left( \delta
\Delta_{h_1}(0)c_{1\uparrow}^{\dag}c_{1\downarrow}^{\dag}
h.c.\right)+\sum_k[c_1\leftrightarrow c_2]\nonumber\\
&& +\sum_k[c_1\leftrightarrow f_1]+\sum_k[c_1\leftrightarrow
f_2]\nonumber\\
&&+\delta\rho(0)\sum \left(c_1^{\dag}c_1\, +\, ...\,+\,
f_2^{\dag}f_2\right) \eea where  $\delta \Delta_i \equiv \delta
\Delta_i (q, \Omega) e^{i(\Omega t - {\bf q}\cdot{\bf r})}$ and
$\Delta \rho \equiv \delta \rho (q, \Omega) e^{i(\Omega t - {\bf
q}\cdot{\bf r})},$ compute fully renormalized $\delta \Delta$ and
$\delta \rho$, and obtain collective modes as the poles of the
generalized susceptibility. Alternatively, the collective modes
can be computed by extending HS approach to finite $q$ and
$\Omega$, see Refs. \onlinecite{LR1,LB,LB2}.

The field $\delta \rho (q, \Omega) \equiv
\delta \rho$ is real, while $\delta \Delta_i (q, \Omega)$ is
generally complex and it is instructive to split it into real and
imaginary parts: $\delta \Delta_j (q, \Omega) = \delta_j^R + i
\delta^I_j$. If the equilibrium gap $\Delta_j$ is real,
$\delta^R_j$ and $\delta^I_j$ describe amplitude (longitudinal)
and phase (transverse) fluctuations of the gap. If the equilibrium
gap is complex, each of $\delta^R_j$ and $\delta^I_j$ describes
amplitude and phase fluctuations. In particular, if in equilibrium
$\Delta_{h1} =\Delta e^{i\phi/2}, \Delta_{h2} =\Delta
e^{-i\phi/2}, \Delta_{e} =-\gamma \Delta$, the relation between
$\delta^R_j$, $\delta^I_j$ and the changes of the amplitudes and
the phases of the three gaps $\left[|\Delta_{h_1}| \to \Delta +
m_{h_1}, \phi/2 \to \phi/2 + \phi_{h_1};\right.$\\
$ \Delta_{h_2} \to \Delta + m_{h_2}, -\phi/2 \to -\phi/2 +
\phi_{h_2};$\\ $\left. |\Delta_{e}| \to -\gamma\Delta + m_{e}, 0
\to \phi_e\right]$ is

\begin{widetext}
\bea\label{eq:delta mf} \left(
\begin{array}{c}
\delta^R_{h_1}\\
\delta^R_{h_2}\\
\delta^R_{e}\\
\delta^I_{h_1}\\
\delta^I_{h_2}\\
\delta^I_{e}\\
\end{array}
\right)&=& \left(
\begin{array}{cccccc}
\cos\frac{\phi}{2}&0&0&-\sin\frac{\phi}{2}&0&0\\
0&\cos\frac{\phi}{2}&0&0&\sin\frac{\phi}{2}&0\\
0&0&-1&0&0&0\\
\sin\frac{\phi}{2}&0&0&\cos\frac{\phi}{2}&0&0\\
0&-\sin\frac{\phi}{2}&0&0&\cos\frac{\phi}{2}&0\\
0&0&0&0&0&-\gamma\\
\end{array}
\right) \left(
\begin{array}{c}
m_{h_1}\\
m_{h_2}\\
m_{e}\\
\Delta \cdot \phi_{h_1}\\
\Delta\cdot \phi_{h_2}\\
\Delta\cdot \phi_{e}\\
\end{array}
\right) \eea
\end{widetext}

Each of the bare vertices gets renormalized by the pairing
interactions and long-range Coulomb interaction.  At weak
coupling, only ladder-type particle-particle renormalizations and
small $q$ particle-hole renormalizations are relevant.  Collecting
the relevant diagrams (see Fig. \ref{fig:fig2b} in the main text),
we obtain the set of coupled equations for fully renormalized
vertices $\overline{\delta\Delta}_i = {\bar \delta}_i^R + i {\bar
\delta}^I_i$ and ${\overline{\delta \rho}} ={\bar \delta}_\rho$ as
 we said in the main text.

The seven branches of collective excitations are obtained from the
condition that ${\text {Det}} {\underline K}(q, \Omega) =0$. Two
of these branches are fluctuations of the overall phase and of the
total density, the others are three longitudinal gap fluctuations
and two different fluctuations of the relative phases of the three
gaps. Some of these fluctuations decouple from the others, but
some are coupled.

The components of $\Pi^{a\,b}_{ii} (q, \Omega)$ can be represented
in the Nambu formalism as \bea\label{eq:LR}
&&\Pi^{a\,b}_{i} (q, \Omega) = \nonumber \\
&& \frac{1}{N_0}T\sum_\omega \int \frac{d^2 k}{2 \pi)^2} Tr\left[
\mathcal{G}_i (k, \omega) \sigma^a \mathcal{G}_i (k+q, \omega +
\Omega) \sigma^b\right] \nonumber\\\eea where $\omega$ is the
fermionic Matsubara frequency,  $\sigma^i$ are the Pauli matrices,
and \bea \mathcal{G}_i (k, \omega)&=&\left(
\begin{array}{cc}
G_{i} (k, \omega)&-F_{i}(k, \omega)\\
-F^{\dag}_{i} (k, \omega)&\tilde G _{i} (k, \omega)
\end{array}
\right)
\eea
where
\bea
{\tilde G}_{i} (k, \omega)&=&-\frac{i\omega+\e_{k,i}}{\omega^2+E^2_i}\nonumber\\
G_{\downarrow\downarrow}&=&-\frac{i\omega-\e_{k,i}}{\omega^2+E^2_i}\nonumber\\
F_{\downarrow\uparrow}&=&-\frac{\Delta_i}{\omega^2+E^2_i}\nonumber\\
F_{\uparrow\downarrow}^{\dag}&=&-\frac{\Delta^*_i}{\omega^2+E^2_i}
\eea Evaluating the integrals, we find that 21 components of
${\underline \Pi}$ are non-zero.

To properly describe all collective excitations, one should keep
the frequency to be of order $\Delta$, as some of the modes exist
only as resonances at $\Omega > 2 \Delta$.  Our goal, however, is
more focused as we are only interested in the 2D plasmon mode and
in the modes which soften at the boundaries of TRSB state. These
modes are the solutions of ${\text {Det}} {\underline K}(q,
\Omega) =0$ at small $\Omega$, and to get these modes one can
safely expand in both $v_F q/\Delta$ and in $\Omega/\Delta$

Evaluating the integrals and converting from Matsubara to real
frequency axis  we obtain the expressions for $\Pi_{ii}^{jk}$ and ${\underline K} (q, \Omega)$ at small $\Omega$ and $\vec{q}$,
 which we presented in Eqs. (\ref{eq:pi_full}) and (\ref{eq:kernel}) in the main text.

Solving for ${\text {Det}} {\underline K} (q, \Omega) =0$, we
obtain seven branches of collective excitations, which we discuss
in the main text.  One can show quite generally that fluctuations
of the overall phase and of the total density are coupled to each
other  but decoupled from other five branches of collective
excitations. One of coupled oscillation of the overall phase and
the total  density is a plasmon mode (see the main text). Among
the other five modes, longitudinal and transverse fluctuations
decouple in ++ and $+-$ phases, but couple in the TRSB state. This
coupling leads to a peculiar structure of low-energy collective
excitations near the boundaries of the TRSB state. We present the
results in the main text.

\subsection{Plasmon mode in a 3D superconductor}

For completeness, we also present the diagrammatic derivation of
the dispersion of a plasmon mode (a coupled oscillation of a phase
of a superconductor order parameter and an electron density) in a
3D superconductor. In 3D,  plasmon frequency tends to a finite
value at $q \to 0$, and the approximation $\Omega \ll \Delta$,
which we used in the previous subsection, is not applicable, at least in the clean limit.

In the dirty limit, the plasmon frequency is small (it can be much
smaller than $\Delta$). A general gradient expansion analysis in
this case shows~\cite{dirty} that the plasma frequency scales with
the density of superconducting electrons (the ``superfluid
density"). In a clean limit, superfluid density coincides with the
full density, and it is reasonable to expect that the plasma
frequency remains the same as in the normal state.

That the plasma frequency is not renormalized in the clean limit
and at $T=0$ has been argued by Anderson back in 1958 on general
grounds (Ref.\cite{and}) and has been shown explicitly by Ohashi
and Takada using an RPA formalism, extended to a superconduting
state~\cite{japa_1}. We reproduce this result in a direct
diagrammatic approach, similar to the one we used in the main text
for the 2D case. For briefness we consider the case of a
single-band s-wave superconductor. The extension to multi-band
systems is straightforward.

\begin{figure}[htp]
\includegraphics[width=0.7\columnwidth]{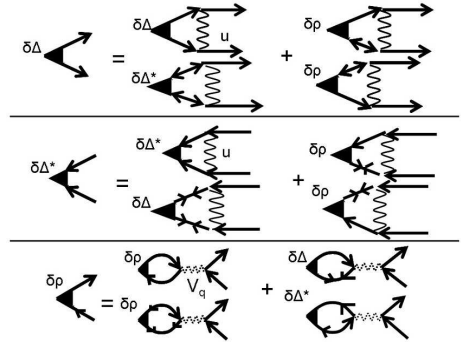}
\caption{\label{fig:plasmon} Diagrammatic representation of the
coupled equations for fluctuations of the total density
$\delta\rho$ and the SC order parameter $\delta\Delta$ (and
$\delta\Delta^*$) for the one band case. The solid and  dotted
wavy lines represent the pairing interaction $u<0$ and unscreened
Coulomb interaction $V_q$. The lines with single and double arrows
represent the normal(G) and anomalous(F) Green functions. The
coupling is due to GF terms which are non-zero when
$\vec{q},\Omega\neq 0$}
\end{figure}

We follow the same strategy as in the main text -- introduce  bare
particle-particle and particle-hole vertices, which correspond to
small variations of a superconducting gap and a total density
($\delta \Delta = \delta^R + i \delta^I$ and $\delta \rho$,
respectively), and express the full vertices in terms of the bare
ones,  using dimensionless $u<0$ for the pairing interaction and
$V_q = A_3/q^2$ for Coulomb interaction in 3D, with $A_3 = 4\pi e^2$.
 The diagrams for the vertices are shown in Fig. \ref{fig:plasmon}

Like in the previous section, we introduce the vector ${\bf
\delta}$ with the components $\delta^R, -\delta^I$, and $\delta
\rho$,
 and write the full vertex ${\bar \delta}$ in the same was as in (\ref{ch_4}), but now with
\bea \label{eq:kernel2}
{\underline K} (q, \Omega)  = \left(
\begin{array}{ccc}
-\frac{2}{u}+\Pi^{11}&0&0\\
0&-\frac{2}{u}+\Pi^{22}& -\Pi^{23}\\
0&-\Pi^{32}& -\frac{1}{N_0V_q}+\Pi^{33}\\
\end{array}
\right) \eea The zeros indicate that the magnitude fluctuations
$\delta^R$ do not couple to the phase and density fluctuations
($\delta^I$ and $\delta \rho$ terms).  The last two fluctuations,
however, couple to each other. The dispersion of the collective
modes are again obtained from the condition ${\text {Det}}
{\underline K} (q, \Omega) =0$. The mode which corresponds to
coupled phase-density oscillations is obtained from
\bea\label{eq:plasmon}
\left(\frac{2}{u}-\Pi^{22}\right)\left(\frac{1}{N_0V_q}-\Pi^{33}\right)=\Pi^{23}\Pi^{32}\eea
Expanding only in $\vec{q}$, we get
\bea\label{eq: forms} \Pi^{23} &=& \frac{i\Omega}{2\Delta}\left[
\mathcal{I}_{\Omega}
+\left(\frac{Q}{2\Delta}\right)^2 \mathcal{I}^{23}_{\Omega}\right]\nonumber\\
\Pi^{32} &=& -\Pi^{23} \nonumber \\
\Pi^{22} &=& \frac{2}{u}-\left(\frac{\Omega}{2\Delta}\right)^2
\mathcal{I}_{\Omega}
+\left(\frac{Q}{2\Delta}\right)^2\mathcal{I}^{22}_{\Omega}\nonumber\\
\Pi^{33} &=& -
\mathcal{I}_{\Omega}-\left(\frac{Q}{2\Delta}\right)^2 \left[
\mathcal{I}^{23}_{\Omega}+\mathcal{I}^{33}_{\Omega}\right] \eea
where $E = \sqrt{\epsilon^2 + \Delta^2}$, $\Lambda$ is the upper
cutoff, and $Q^2=\langle (\vec{v}_F.\vec{q})^2\rangle =
\frac{v_F^2q^2}{3}$ in 3D (and $\frac{v_F^2q^2}{2}$ in 2D).
In $\Pi^{22}$ we have used the BCS gap equation that tells us \be
-\frac{2}{u} = \int_{\-\Lambda}^\Lambda \frac{d\epsilon}{E}
\label{ch_5} \ee
Also, \bea\label{eq:I} \mathcal{I}_{\Omega}&=&\int
\frac{\Delta^2}{E\left(E^2-\frac{\Omega^2}{4}\right)}\nonumber\\
\mathcal{I}_{\Omega}^{22}&=&\int
\frac{\Delta^4\left(3E^2-\frac{\Omega^2}{4}\right)}{2E^3\left(E^2-\frac{\Omega^2}{4}\right)^2}\nonumber\\
\mathcal{I}_{\Omega}^{23}&=&\int
\frac{\Delta^4\left(E^2\left(2E^2-5\Delta^2\right)+\left(2E^2-3\Delta^2\right)\left(\frac{\Omega}{2}\right)^2\right)}{2E^5\left(E^2-\frac{\Omega^2}{4}\right)^2}\nonumber\\
\mathcal{I}_{\Omega}^{33}&=&\int
\frac{\Delta^4}{E^3\left(E^2-\frac{\Omega^2}{4}\right)}\nonumber\\
\eea

Eq. \ref{eq:plasmon} now becomes, to the leading order in $q$:
\be\label{eq:plasmon2} \Omega^2 =N_0 V_q
Q^2\left[\mathcal{I}_{\Omega}^{22}-\left(\frac{\Omega}{2\Delta}\right)^2\left(\mathcal{I}_{\Omega}^{33}-\mathcal{I}_{\Omega}^{23}\right)\right]+O(Q^2)\ee
Using
\bea\label{eq:eval}
\mathcal{I}_{\Omega}^{22}&=&2+\left(\frac{\Omega}{2\Delta}\right)^2\int
\frac{\Delta^4\left(5E^2-2\left(\frac{\Omega}{2}\right)^2\right)}{2E^5\left(E^2-\frac{\Omega^2}{4}\right)^2}\nonumber\\
\mathcal{I}_{\Omega}^{33}-\mathcal{I}_{\Omega}^{23}&=&\int
\frac{\Delta^4\left(5E^2-2\left(\frac{\Omega}{2}\right)^2\right)}{2E^5\left(E^2-\frac{\Omega^2}{4}\right)^2}
\eea we immediately find that \bea\label{eq:eval2}
\mathcal{I}_{\Omega}^{22}-\left(\frac{\Omega}{2\Delta}\right)^2\left(\mathcal{I}_{\Omega}^{33}-\mathcal{I}_{\Omega}^{23}\right)&=&2
\eea and hence

\be\label{eq:plasmon3} \Omega^2=2N_0 V_q Q^2\ee which is the same
result as in the normal state. Substituting the expressions for
$V_q = 4\pi e^2/q^2$, $Q^2 = v^2_F q^2/3$, $N_0 = m p_F/(2\pi^2)$,
and using the relation between $p_F$ and the density of fermions
$p^3_F/(3\pi^2) = n$, we obtain \be\label{eq:plasmon4}
\Omega^2=\frac{4\pi n e^2}{m} = \Omega^{2}_{pl}\ee which is the
same plasma frequency as in the normal state. This result is
well-known starting from the Anderson work~\cite{and}. Like we
said, our goal was just to demonstrate how this result can be
re-derived in a direct diagrammatic approach.

At a finite $T\leq T_c$ and/or in the presence of impurity
scattering, coupled density and phase fluctuations are more
complex, and near $T_c$ there exists a weakly damped, near-gapless
Carlson-Goldman mode~\cite{cg}. The evolution of plasma
oscillations with increasing $T$ and/or impurity scattering are
not fully understood as only the cases $\Omega = \Omega_{pl}$  and
$\Omega \ll \Delta \ll \Omega_{pl}$ have been analyzed in detail
(see, e.g., Ref. \onlinecite{LR1}).  The diagrammatic approach
which we present here offers the way to obtain the results for all
$T$ and also with and without impurity scattering.

\end{document}